\documentclass[twocolumn]{aastex62}

\shortauthors{Blancato et al.}

\usepackage[T1]{fontenc}
\usepackage{ae,aecompl}
\usepackage{lipsum}
\usepackage[]{algorithm2e}

\usepackage{graphicx}
\usepackage{amsmath}
\usepackage{amssymb}
\usepackage{enumitem}

\newcommand{\RN}[1]{%
\textup{\uppercase\expandafter{\romannumeral#1}}
}

\usepackage{hyperref}

\definecolor{mycolor}{HTML}{045a8d}

\hypersetup{
colorlinks,
citecolor=mycolor,
filecolor=mycolor,
linkcolor=mycolor,
urlcolor=mycolor
}

\hypersetup{urlcolor=mycolor}
\usepackage{ulem}
\shorttitle{[Mg/Si] variations}

\begin{document}

\title{\uppercase{Variations in $\alpha$-element ratios trace the chemical evolution of the disk}}
\correspondingauthor{Kirsten Blancato}
\email{knb2128@columbia.edu}

\author{Kirsten Blancato}
\affil{\textit{Department of Astronomy, Columbia University, 550 West 120th Street, New York, NY, 10027, USA}}

\author{Melissa Ness}
\affiliation{\textit{Department of Astronomy, Columbia University, 550 West 120th Street, New York, NY, 10027, USA}}
\affiliation{\textit{Center for Computational Astrophysics, Flatiron Institute, 162 Fifth Avenue, New York, NY, 10010, USA}}

\author{Kathryn V. Johnston}
\affiliation{\textit{Department of Astronomy, Columbia University, 550 West 120th Street, New York, NY, 10027, USA}}
\affiliation{\textit{Center for Computational Astrophysics, Flatiron Institute, 162 Fifth Avenue, New York, NY, 10010, USA}}

\author{Jan Rybizki}
\affiliation{\textit{Max-Planck-Institut f\"{u}r Astronomie, K\"{o}nigstuhl 17, D-69117 Heidelberg, Germany}}

\author{Megan Bedell}
\affiliation{\textit{Center for Computational Astrophysics, Flatiron Institute, 162 Fifth Avenue, New York, NY, 10010, USA}}

\vspace{10px}
\begin{abstract}
It is well established that the chemical structure of the Milky Way exhibits a bimodality with respect to the $\alpha$-enhancement of stars at a given [Fe/H]. This has been studied largely based on a bulk $\alpha$ abundance, computed as a summary of several individual $\alpha$-elements. Inspired by the expected subtle differences in their nucleosynthetic origins, here we probe the higher level of granularity encoded in the inter-family [Mg/Si] abundance ratio. Using a large sample of stars with \texttt{APOGEE} abundance measurements, we first demonstrate that there is additional information in this ratio beyond what is already apparent in [$\alpha$/Fe] and [Fe/H] alone. We then consider \textit{Gaia} astrometry and stellar age estimates to empirically characterize the relationships between [Mg/Si] and various stellar properties. We find small but significant trends between this ratio and $\alpha$-enhancement, age, [Fe/H], location in the Galaxy, and orbital actions. To connect these observed [Mg/Si] variations to a physical origin, we attempt to predict the Mg and Si abundances of stars with the galactic chemical evolution model \textit{Chempy}. We find that we are unable to reproduce abundances for the stars that we fit, which highlights tensions between the yield tables, the chemical evolution model, and the data. We conclude that a more data-driven approach to nucleosynthetic yield tables and chemical evolution modeling is necessary to maximize insights from large spectroscopic surveys.
\end{abstract}


\section{Introduction}
\label{sec:introduction}
Alpha ($\alpha$) elements (such as Mg, Ti, Si, Ca) are primarily produced through the successive fusion of helium nuclei in high-mass ($M_{*}$ \textgreater~8 $M_{\odot}$) stars, and are released to the interstellar medium (ISM) when these stars explode as core collapse supernovae (CC-SN). In contrast, iron-peak elements (such as Fe, Mn, Cr, Ni) are produced in both CC-SN and SN Ia supernovae (SN Ia). The time delay between production of yields from CC-SN and SN Ia leads to informative contrasts between different families of elements. For example, the relative abundance of $\alpha$-elements to iron (Fe) has been of longstanding interest as an indicator of the star-formation history of a galaxy, as well as the contribution of yields from CC-SN versus SN Ia at the site of star-formation \citep{venn, tinsley79, pagel}. 

Fundamental chemical properties of the Milky Way have been revealed by investigating $\alpha$ and Fe abundances. In particular, it has become well established that the chemical structure of the Galaxy exhibits a bimodality with respect to the $\alpha$-enhancement of stars at a given [Fe/H]. This bimodality was first observed locally in the solar neighborhood \citep{fuhrmann98, prochaska00, reddy06, adibekyan12, bensby14}, and was also apparent within the first year of the  Apache Point Observatory Galactic Evolution Experiment (\texttt{APOGEE}) survey, where stars within the solar circle ($d$ \textless ~1 kpc) were observed to follow two ``sequences" in the [$\alpha$/Fe] vs. [Fe/H] plane. As shown in \cite{anders14}, the ``low-$\alpha$ sequence" is focused near solar [$\alpha$/Fe] spanning a range of metallicities from [Fe/H] $\sim$-0.8 to 0.4 dex, while the ``high-$\alpha$ sequence" covers a range of enriched $\alpha$ abundances, from $\sim$0.3 dex at [Fe/H] $\sim$-1.0 to $\sim$0.1 dex at [Fe/H] $\sim$ 0. At [Fe/H] $\sim$ 0.1, the high- and low-$\alpha$ sequences appear to merge. 

Beyond the solar neighborhood, large surveys such as \texttt{APOGEE} and \textit{Gaia} have enabled the empirical characterization of the bimodal $\alpha$ sequence throughout the Milky Way disk \citep{bovy12, nidever14, hayden15, mackereth17}. For instance, using SDSS/SEGUE spectra \cite{bovy12} examine how the $\alpha$ abundances of stars vary with location from disk midplane ($|z|$ $\sim$ 0.3 - 3 kpc), as well as with Galactocentric radius (R = 5 - 12 kpc). They find that, compared to stars with low-$\alpha$ abundances, the population of stars with the highest [$\alpha$/Fe] enrichment are more vertically extended, yet radially concentrated. Using a larger sample of stars from \texttt{APOGEE}, \cite{hayden15} confirm this result and further characterize how the ratio of stars with low versus high-$\alpha$ abundances varies throughout the disk. They find that the inner disk (R $\lesssim$ 9 kpc) is comprised of both low- and high-$\alpha$ sequence stars; however, low-$\alpha$ sequence stars are primarily confined close to the disk midplane ($|z|$ $\lesssim$ 1 kpc) and high-$\alpha$ sequence stars are primarily located at $|z|$ $\sim$ 0.5 - 2 kpc. Contrary to the inner disk, at all distances from the midplane the outer disk (R $\gtrsim$ 9 kpc) is markedly devoid of stars with high-$\alpha$ abundances. This observed chemical structure of the Milky Way has been cited as evidence for ``inside-out" \citep{inside-out} and ``upside-down" \citep{upside-down} formation scenarios of the disk components of our Galaxy where, radially, the central disk was formed before the outer disk and, vertically, the thick disk was formed before the thin disk. 

Although the $\alpha$-enriched component of the Milky Way has come to be associated with a ``hotter disk" and the solar-$\alpha$ component often associated with a ``cooler disk" \citep{bensby03, navarro11, bovy12}, the origin of these two populations, and whether they are unique, is still debated \citep{haywood, toy, bovy12a}. Note that these hotter and cooler populations do not necessarily follow the same morphology as have previously been identified as the ``thick" and ``thin" disks \citep{gilmore83}. However, there has been growing evidence that, in addition to their chemical differences, the two sequences are also dynamically distinct. For example, using a sample of stars from \texttt{APOGEE} and \textit{Gaia}, \cite{mac19} find that at fixed age and [Fe/H], the low- and high-$\alpha$ sequences display different age-velocity dispersion relationships with respect to both radial and vertical velocity dispersion. As discussed in \cite{mac19}, these kinematic differences between the low- and high-$\alpha$ sequence are suggestive of disparate heating mechanisms contributing to the formation of the two populations. Similarly, \cite{gandhi} report differences between the low- and high-$\alpha$ sequence in terms of their orbital actions ($J_{\phi}$, $J_{R}$, $J_{z}$) at all stellar ages. 

While differences between the high- and low-$\alpha$ sequence have been characterized empirically, the origin of the bimodality remains an open question. Nonetheless, recent simulation work modeling Milky Way-like galaxies has made substantial progress \citep{grand18, mac19, clarke19}. For example, analyzing the large-volume EAGLE cosmological simulation, \cite{mack18} find that a bimodal $\alpha$ sequence occurs in galaxies that experience an early phase of anomalously rapid mass accretion. Consequently, they report that only $\sim$5\% of Milky Way-like galaxies in the EAGLE volume exhibit a bimodal $\alpha$ sequence similar to our Galaxy's, implying that this chemical structure is rare. More recently, \cite{clarke19} propose that clumpy star-formation is responsible for the observed bimodal $\alpha$ sequence. Using \texttt{GASOLINE} to perform high-resolution simulations, \cite{clarke19} reproduce the Milky Way's chemical bimodality in the [O/Fe]-[Fe/H] plane. Investigating the birth sites of stars in the two sequences, they find that stars in the high-$\alpha$ sequence are formed in clumps that start off with low-$\alpha$ abundances, but rapidly self-enrich due to high star formation rates (SFRs). Meanwhile, low-$\alpha$ sequence stars are the product of a more extended star-formation mode that occurs with a substantially lower SFR. Since the incidence of clumps are common in high-redshift galaxies, \cite{clarke19} conclude that chemical bimodality should be prevalent among Milky Way-mass galaxies. 

Typically, the $\alpha$ abundances investigated in observational and theoretical studies are computed as an average of many individual $\alpha$-elements. However, there is additional information in the relative abundance of different $\alpha$-elements themselves. We know from stellar nucleosynthesis theory that the different $\alpha$-elements vary in the details of their production mechanisms. Given the precision of recent spectroscopic surveys like \texttt{APOGEE}, \texttt{GALAH}, and \texttt{LAMOST}, and the many abundance measurements now available for hundreds of thousands of stars, we can begin to go beyond considering a mean $\alpha$ abundance and examine the information encoded by individual $\alpha$-elements. Full exploitation of the information contained in these multi-element abundance vectors, and how this vector varies with dynamical properties, is still underway. Recently, \cite{weinberg18} has mapped multiple \texttt{APOGEE} abundances from R = 3 - 15 kpc and $|z|$ = 0 - 2 kpc. This includes the comparison of various elements to Mg, including $\alpha$-elements (like S, Si, O, and Ca), light odd-$Z$ elements (like Al, P, K, and Z), as well as iron-peak elements (like Cr, Mn, Fe, V, Co, and Ni). 

Through this broad exploration, they find small variations among the different abundance ratios throughout the disk. By considering these subtle differences between inter-family element combinations in depth, more details of the Galaxy's formation history, and the physics of nucleosynthesis, can be gleaned.

For example, a detailed examination of an inter-family abundance ratio has been carried out for the Sagittarius dwarf galaxy, where stars are observed to be deficient in magnesium (Mg) compared to silicon (Si) \citep{mcwilliam13, hassel17, carlin18}. The discrepancy between the enrichment of Mg and Si has been interpreted as suggesting the Sagittarius galaxy was formed with a ``top-light" stellar initial mass function (IMF), meaning an IMF with fewer high-mass stars compared to the canonical IMF. This is argued to be a consequence of varying yield dependencies on stellar mass between $\alpha$-elements like Mg and O and $\alpha$-elements like Si and Ca. However, this argument is complicated by the fact that some $\alpha$-elements (like Si) are also produced by SN Ia \citep{tinsley79}. 

The interpretation of Sagittarius [Mg/Si] observations are supported by theoretically motivated expected differences in how various $\alpha$-elements are produced. As discussed in \cite{hassel17}, hydrostatic $\alpha$-elements, such as Mg and O, are produced in massive stars during the hydrostatic burning phase, and these elements get ejected into the ISM during CC-SN explosions. In constrast, explosive $\alpha$-elements, such as Si and Ca, are produced in massive stars during the explosive nucleosynthesis leading up to the CC-SN explosion. While both hydrostatic and explosive $\alpha$-elements are produced in massive stars and released via CC-SN, explosive $\alpha$-elements are produced in shells that lie closer to the cores of massive stars, whereas hydrostatic $\alpha$-elements are produced in the outermost shells. This makes the yields of hydrostatic $\alpha$-elements more dependent on the mass of the star, while explosive $\alpha$-element yields are relatively independent of stellar mass \citep{woosley95}. In this regard, different $\alpha$-elements can be used to probe the population of massive stars at a given epoch of star-formation. 

Inspired by the Sagittarius results, as well as the increasing precision and wealth of multi-abundance measurements of stars located throughout the Milky Way's disk, in this paper we perform a focused investigation of the magnesium to silicon abundance ratio. By considering this specific inter-family element ratio, we are able to attain a more resolved understanding of how $\alpha$-elements vary in the Galaxy. We are also able to isolate particular enrichment events associated only with the production of these inter-family elements. We do this by examining how the ratio of Mg to Si varies with age, as well as dynamics, and find that the low- and high-$\alpha$ sequences exhibit markedly different behavior. To put these results in the context of the Galaxy's chemical evolution, we further attempt chemical evolution modeling to extract both ISM and stellar population parameters at the time of star-formation. While in this study we concentrate only on the ratio of Mg to Si, our methods are generalizable. An in-depth analysis of many inter-family abundance ratios, and even a full matrix of element ratios, appears to be a promising avenue for extracting the most information from large spectroscopic surveys. These detailed characterizations of particular stellar abundance ratios (including [Mg/Si]) are also of interest to studies of planet formation and occurrence \citep{adibekyan15}.

This paper is organized as follows. In Section \ref{sec:theory} we use yield tables to explore theoretical Mg and Si contributions from different nucleosynthetic sources. Then in Section \ref{sec:data} we introduce the data and methods of the paper including: the \texttt{APOGEE} sub-sample, clustering of the low- and high-$\alpha$ sequences, and empirical motivation for examining the ratio of Mg to Si. In Section \ref{sec:results} we present the main empirical results of the paper, showing how [Mg/Si] varies: between the low- and high-$\alpha$ sequences, with age and metallicity, throughout the Galactic disk, and with orbital parameters. In the second half of the paper we focus on interpreting these empirical results within the context of Galactic chemical evolution (GCE). To do this, in Section \ref{sec:chempy} we use \textit{Chempy} to fit chemical evolution models to a sample of \texttt{APOGEE} stars and examine variations with the IMF slope, number of SN Ia, and ISM parameters. In this section we discuss the limitations we encounter when attempting to fit these models to a diverse set of stars. Finally, in Section \ref{sec:conclusion} we distill the main takeaways of both the empirical results and attempted chemical evolution modeling, and discuss potential paths forward.


\section{Expected Mg and Si yields}
\label{sec:theory}
Before empirically characterizing the ratio of magnesium to silicon throughout the Milky Way, we further motivate examining this abundance ratio by exploring theoretical Mg and Si yields as expected from yield tables. We do this using the GCE code \textit{Chempy} \citep{chempy}, which is discussed in more detail in Section \ref{sec:chempy}. In short, using \textit{Chempy} we initialize a simple stellar population (SSP) with a Chabrier IMF \citep{chabrier03} and evolve the SSP from 0 to 13.5 Gyr in 1350 linear-spaced steps, keeping track of enrichment from CC-SN, SN Ia, and asymptotic giant branch (AGB) stars. The yields from CC-SN and AGB stars both depend on the mass of dying stars at each time step, whereas yields from SN Ia, parameterized as a power-law delay time distribution (DTD) \citep{maoz10}, are independent of stellar mass. The yields we report throughout this paper are the net yields, which is the newly synthesized material from these nucleosythetic channels. 

We consider two initial SSP metallicities. This includes an SSP with a solar-like metallicity ($Z$ = 0.01), as well as a metal-poor SSP with $Z$ = 0.0001. We also compute SSPs assuming two sets of yield tables. The first is the \textit{Chempy} default yield tables which include CC-SN yields from \cite{nomoto13}, SN Ia yields from \cite{seitenzahl13}, and AGB yields from \cite{karakas10}. The alternative \textit{Chempy} yield set instead uses CC-SN yields from \cite{chieffi04}, SN Ia yields from \cite{thielemann03}, and AGB yields from \cite{ventura13}. In the following figures we only show the Mg and Si yields from the default yield set at $Z$ = 0.0001 and $Z$ = 0.01. The alternative yield set exhibits similar trends as the default yield set; however, for the alternative yield set there is less of a difference in the yields produced at the two metallicities.

\begin{figure}[tp]
\centering
\includegraphics[width=1\columnwidth]{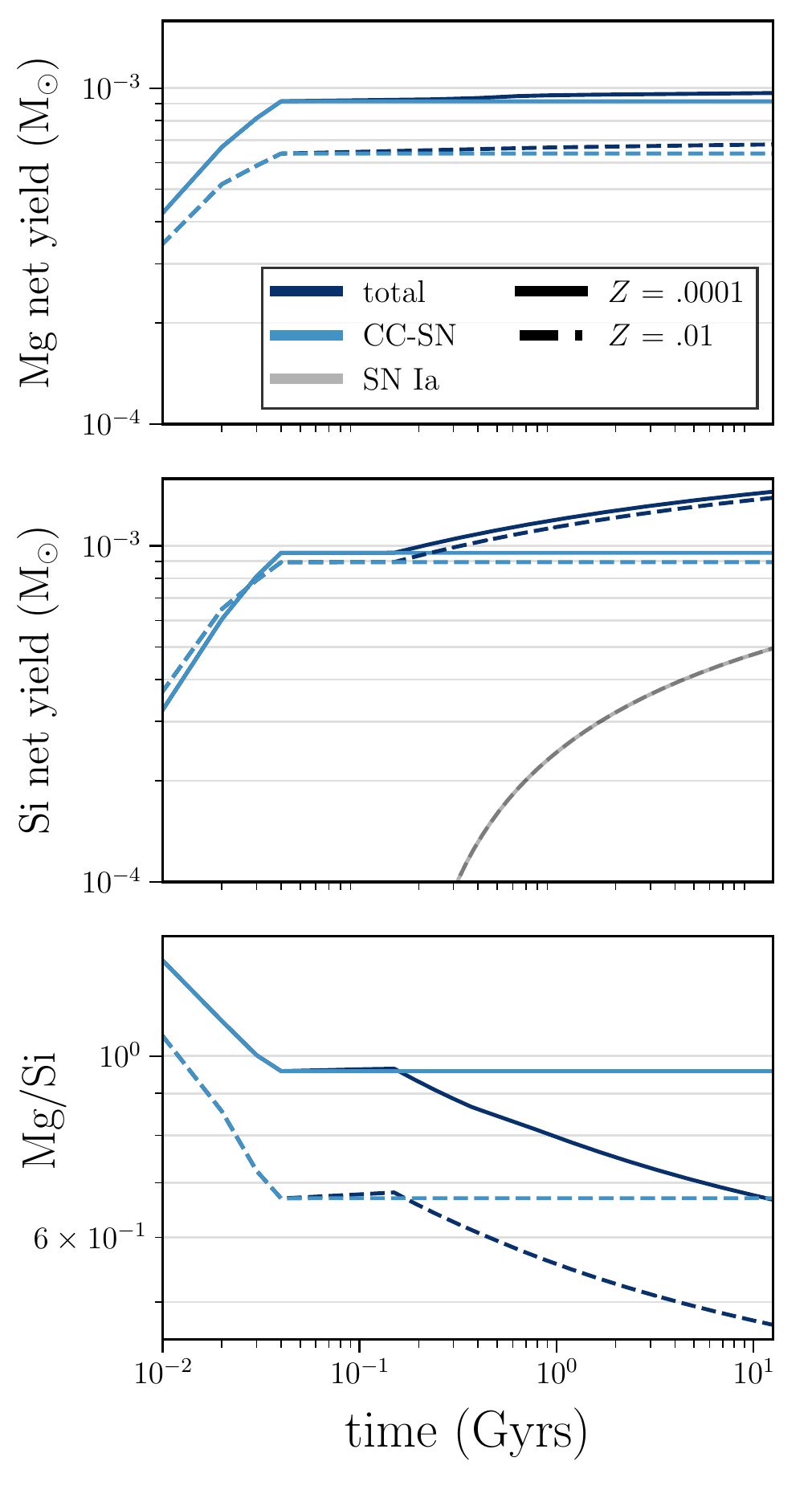}
\caption{\textbf{\textit{Upper two panels:}} expected magnesium (Mg) and (Si) net yields for a 1 M$_{\odot}$ SSP assuming a Chabrier IMF and metallicities of $Z$ = 0.0001 (dashed) and $Z$ = 0.01 (solid). At each time step, the cumulative net yields from CC-SN \citep{nomoto13} and SN Ia \citep{seitenzahl13} are shown. The AGB yields (not shown) are negligible compared to the SN yields. \textbf{\textit{Bottom panel:}} expected ratio of Mg to Si net yields throughout time from the same SSPs. The Si produced in SN Ia reduces the total Mg/Si ratio after 100 Myr.}
\label{fig:theory_mgsi}
\end{figure}

\begin{figure}[tp]
\centering
\includegraphics[width=1\columnwidth]{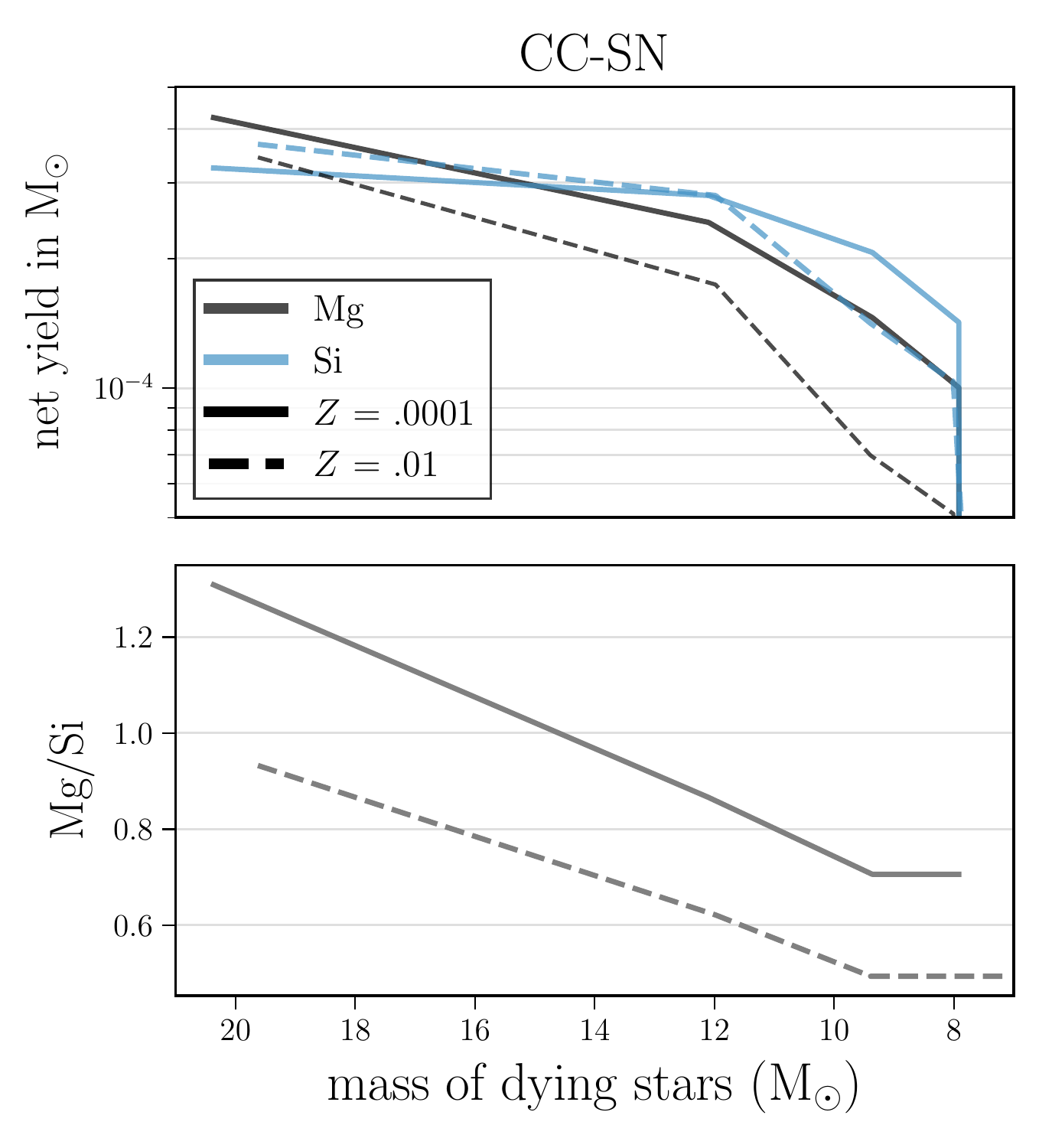}
\caption{\textbf{\textit{Top panel:}} expected magnesium and silicon yields from CC-SN assuming the same SSP parameters and yield tables as in Figure \ref{fig:theory_mgsi}. Shown are the yields generated from dying stars with masses from $M_{*}$ = 8 - 20 M$_{\odot}$. The yields of Si are more constant with stellar mass than those of Mg, especially for the $Z$ = 0.0001 SSP. \textbf{\textit{Bottom panel:}} expected ratio of Mg to Si net yields as a function stellar mass for both the $Z$ = 0.1 (dashed) and $Z$ = 0.0001 (solid) SSPs. As stellar mass decreases, more Si is produced compared to Mg.}
\label{fig:ccsn}
\end{figure}

The \textit{\textit{upper two panels}} of Figure \ref{fig:theory_mgsi} show the cumulative Mg and Si yields as a function of time. The total net yields summing the contributions from the three nucleosynthetic channels, as well as the individual yields from CC-SN and SN Ia, are shown. The AGB yields are negligible compared to the SN yields and scale of the plots, and thus are not shown. For both magnesium and silicon, there is an early phase of element production from CC-SN, which ends after $\sim$40 Myr. After this time, both the Mg and Si yields remain constant until $\sim$150 Myr, at which point SN Ia ejecta begins to enrich the ISM. As seen in the figure, SN Ia produce a negligible amount of Mg compared to Si, which results in the Mg yield remaining relatively constant until present day, while the Si yield continues to increase. To more closely examine the differences between the two elements, the \textit{\textit{bottom panel}} of Figure \ref{fig:theory_mgsi} shows how the ratio of Mg to Si evolves over time. Here we see that Mg/Si decreases with time, first due to more Si ejecta from CC-SN, then due to the Si ejecta from SN Ia. While there is a normalization offset and small differences comparing the $Z$ = 0.01 and $Z$ = 0.0001 SSPs, the overall trend is persistent at both metallicities. 

Focusing on the CC-SN yields, Figure \ref{fig:ccsn} shows the Mg and Si yields per mass of dying star. For the metal-poor SSPs, the higher-mass stars ($\textgreater$ 15 M$_{\odot}$) produce more Mg than Si, whereas lower-mass stars ($\textless$ 15 M$_{\odot}$) produce more Si than Mg. This results in an ISM that is Mg/Si rich at early times when the massive stars are exploding, and then becomes Mg/Si deficient once the lower mass CC-SN begin to explode at later times. For the solar-metallicity SSP we see that, while Si and Mg also exhibit different dependencies on stellar mass, the Si yields are greater than the Mg yields at every mass. So while the ratio of Mg to Si yields for the $Z$ = 0.01 SSP also decreases with time, the population is always deficient in Mg compared to Si. 

From this exploration of the magnesium and silicon yields generated throughout the lifetime of a SSP, we confirm the expected behavior of these yields discussed in Section \ref{sec:introduction} and build intuition for how the abundance ratio of [Mg/Si] might vary in the simplest case. Considering the theoretical expectations, Figure \ref{fig:theory_mgsi} shows that the yield tables predict Si, but not Mg, to be produced in SN Ia, which causes Mg/Si to continually decrease once SN Ia enrichment commences. Additionally, Figure \ref{fig:ccsn} confirms that the amount of Mg (hydrostatic) and Si (explosive) generated in CC-SN exhibits different dependencies on stellar mass, making Mg/Si sensitive to the high-mass end slope of the IMF. 

While these trends in Mg to Si are easily understood for the case of a SSP, in reality the interpretation of [Mg/Si] will not be as straightforward. Stars can form from an ISM that is enriched by several generations of stellar populations, at different starting metallicities, and exhibit incomplete mixing, all which will alter the [Mg/Si] abundance. To begin to disentangle what can be learned from variations in the ratio of Mg to Si, these variations must first be empirically quantified. With this goal, in the next section we describe the data we use to carry out the characterization of [Mg/Si] throughout the Milky Way disk.

\section{Data and Methods}
\label{sec:data}

\subsection{Main data sample}
\label{sec:main_sample}
To investigate how the Mg to Si abundance ratio varies in the Milky Way, we use data from the publicly available \texttt{APOGEE} Data Release 14 (DR14) \citep{abolfathi17}. \texttt{APOGEE} DR14 is a main SDSS-IV \citep{blanton17} campaign carried out with the \texttt{APOGEE} near-IR spectrograph \citep{wilson10} on the 2.5 meter SDSS telescope \citep{gunn98} located at Apache Point Observatory. The \texttt{APOGEE} Stellar Parameter and Chemical Abundances Pipeline (ASPCAP) \citep{perez16} is used to derive stellar parameters and chemical abundances by $\chi^{2}$ fitting to 1D local thermal equilibrium models. Produced is a data catalog consisting of stellar atmosphere parameters (e.g. $T_{\rm eff}$, log $g$), as well as chemical abundances including a global metallicity, [M/H], an $\alpha$ abundance, [$\alpha$/M] (which is a combination of O, Mg, Si, S, Ca, and Ti), and measurements of 19 individual elemental abundances: C/C I, N, O, Na, Mg, Al, Si, P, S, K, Ca, Ti/Ti II, V, Cr, Mn, Fe, Co, Ni, and Rb. 

The entire \texttt{APOGEE} DR14 ASPCAP catalog contains 277,371 stars. To obtain stellar age estimates, we match this catalog with the \cite{ness16} catalog, which provides stellar ages for 73,151 \texttt{APOGEE} stars included in DR14. \cite{ness16} derive these ages using \textit{The Cannon} \citep{ness15} and infer stellar masses to a precision of $\sim$0.07 dex based on the \texttt{APOGEE} spectra. These mass estimates translate to stellar age estimates with $\sim$40\% errors, and are found to be primarily based on CN absorption features.

To examine trends with structural and orbital properties, we further match the \texttt{APOGEE} sample with a \textit{Gaia} product catalog. \cite{sanders} include distances, Galactocentric coordinates, and actions for a majority of stars in our sample. Distances are derived with a Bayesian approach using spectroscopic, photometric, and astrometric properties of each star. From these distances, \cite{sanders} compute the Galactocentric radius (R) of each star, as well as the distance from the disk midplane ($z$). The angular momenta ($J_{\phi}$), vertical actions ($J_{z}$), and radial actions ($J_{R}$) are computed in the \cite{mcmillan} potential using the St\"ackel Fudge method \citep{sanders16}.

Finally, we discard stars with \texttt{STAR$\_$BAD} indicated in the \texttt{APOGEE} \texttt{ASPCAPFLAG} and also remove stars without [Fe/H], [Mg/Fe], [Si/Fe], and [$\alpha$/Fe] measurements. This results in 72,125 stars with which to carry out our study. The median (min, max) of the relevant chemical abundances are -0.13 dex (-1.17 \textless~[Fe/H] \textless~0.61 dex) in [Fe/H], 0.034 dex (-0.28 \textless~[$\alpha$/Fe] \textless~0.42 dex) in [$\alpha$/Fe], 0.050 dex (-0.48 \textless~[Mg/Fe] \textless~0.61 dex) in [Mg/Fe], and 0.027 dex (-0.61 \textless~[Si/Fe] \textless~0.59 dex) in [Si/Fe]. The median (min, max) of the stellar atmosphere parameters are 2.46 (1.05 \textless~log $g$ \textless~3.72) in log $g$ and 4780 K (3980 \textless~$T_{\rm eff}$ \textless~5800 K) in $T_{\rm eff}$. The median Galactocentric radius and distance from the disk midplane of the sample are R = 9.47 kpc and |$z$| = 0.37 kpc, and the 3$\sigma$ spatial extent spanned is 0.14 \textless ~R \textless ~44.6 kpc and 0.00 \textless ~|$z$| \textless ~10.7 kpc. Lastly, the median (3$\sigma$ range) of the actions are 2050 kpc km s$^{-1}$ (-390 \textless~$J_{\phi}$ \textless~3570 kpc km s$^{-1}$) in $J_{\phi}$, 1.51 (-0.81 \textless~log($J_{R}$) \textless~2.76) in log($J_{R}$), and 0.94 (-1.67 \textless~log($J_{z}$) \textless~2.50) in log($J_{z}$). 

\subsection{Additional datasets}
We assume that the measured [Mg/Fe] and [Si/Fe] abundances of stars reflect their abundances at the time of birth. While stochastic effects, like binary interactions and planetary engulfment, might change abundances over time, we expect these affects to be negligible. Additionally, given the narrow temperature range of our star sample, we also assume that dust does not differentially impact the measured abundances.

To test that our empirical results are robust, we corroborate our findings with two other samples: a sample of red clump (RC) stars (which are constrained in $T_{\rm eff}$ and log $g$) and a sample of main-sequence stars with chemical abundance measurements from the High Accuracy Radial velocity Planet Searcher (HARPS). In the following sections we describe and motivate the use of each of these datasets.

\subsubsection{\texttt{APOGEE} red clump stars}
\label{sec:rc}
RC stars are low-mass, core helium-burning stars. As discussed in \cite{rc-girardi}, the constancy of the core mass for these $\sim$1.5 $M_{\odot}$ stars at the start of the core helium-burning phase is what leads these stars to ``clump" to the same luminosity in the color-magnitude diagram. For similar reasons, RC stars also span narrower ranges in stellar parameters such as $T_{\mathrm{eff}}$ and log $g$ compared to red giant branch (RGB) stars. This property makes RC stars a good check of our results, and if what we find is dependent on stellar atmosphere parameters. 

The sample of RC stars we use is from the \texttt{APOGEE} Red-Clump (RC) Catalog, which is derived from the main \texttt{APOGEE} stellar catalog based on the procedure developed in \cite{bovyrc}. The selection makes use of both photometric and spectroscopic data, and identifies probable RC stars based on their metallicity, color, effective temperature, and surface gravity. This selection results in minimal contamination from RGB stars. The DR14 RC catalog contains 29,502 RC stars with distances accurate to 5-10\%, and 18,357 of these stars have age estimates from \cite{ness16}. The median value (min, max) of the chemical abundances for the RC sample are -0.13 dex (-0.90 \textless~[Fe/H] \textless~0.51 dex) in [Fe/H], 0.029 dex (-0.14 \textless~[$\alpha$/Fe] \textless~0.40 dex) in [$\alpha$/Fe], 0.038 dex (-0.31 \textless~[Mg/Fe] \textless~0.53 dex) in [Mg/Fe], and 0.032 dex (-0.41 \textless~[Si/Fe] \textless~0.46 dex) in [Si/Fe]. The median (min, max) of the stellar atmosphere parameters are 2.45 (1.80 \textless~log $g$ \textless~3.13) in log $g$ and 4880 K (4190 \textless~$T_{\rm eff}$ \textless~5440 K) in $T_{\rm eff}$. As expected, the RC stars are in a narrower range in log $g$ and $T_{\mathrm{eff}}$ than the main sample described in Section \ref{sec:main_sample}.

\subsubsection{HARPS solar twins}
\label{sec:harps}
Solar twin stars are main-sequence G dwarfs that are spectroscopically similar to the Sun. 
A typical solar twin has an effective temperature within 100 K, surface gravity log $g$ within 0.1 dex, and bulk metallicity or iron abundance [Fe/H] within 0.1 dex of the solar values \citep[e.g.][]{Ramirez2014, Nissen2015}. 
As a result, the spectra of solar twins may be compared differentially to the solar spectrum with minimal reliance on stellar atmospheric models, yielding extremely high-precision (0.01 dex level) abundance measurements \citep{Bedell2014}. 
Similarly precise spectroscopic parameters and therefore isochronal ages are also possible for these stars \citep{Ramirez2014, Spina2018}. 

We use a set of 79 solar twin stars observed at high signal-to-noise and high resolution with the HARPS spectrograph. 
These stars have highly precise ages derived from isochrone fits to their spectroscopic parameters with an estimated uncertainty of 0.4 Gyr in \citet{Spina2018}. 
Using the same spectra and parameters, abundances for several $\alpha$-elements including Mg and Si were derived in \citet{Bedell2018} through a differential equivalent width technique. 

The stars in this dataset are limited in both number and scope: all are located within $\sim100$ pc of the Sun and, by definition, all have approximately solar metallicity. 
Despite these limitations, the precision of age and abundance measurements in this sample makes it a valuable source of information. This sample also serves as a test for the generalizability of our results to main-sequence stars.

\subsection{Soft clustering of low- and high [$\alpha$/Fe] stars}
\label{sec:clustering}
To quantify differences in [Mg/Si] between stars with solar $\alpha$ abundances and those with enriched $\alpha$ abundances, we first separate the full \texttt{APOGEE} sample into a `low-$\alpha$' sequence and a `high-$\alpha$' sequence. We cluster the subsample of \texttt{APOGEE} stars described in Section \ref{sec:main_sample} based on two input features, $\vec{X}$ = $\{$[$\alpha$/Fe], [Fe/H]$\}$, which is the parameter space in which the two sequences are typically identified. Visually, the two proposed clusters in this feature space are distributed anisotropically and not cleanly separable, so we decide to achieve a soft clustering through fitting a mixture model. 

\begin{figure}[tp]
\centering
\includegraphics[width=1.0\columnwidth]{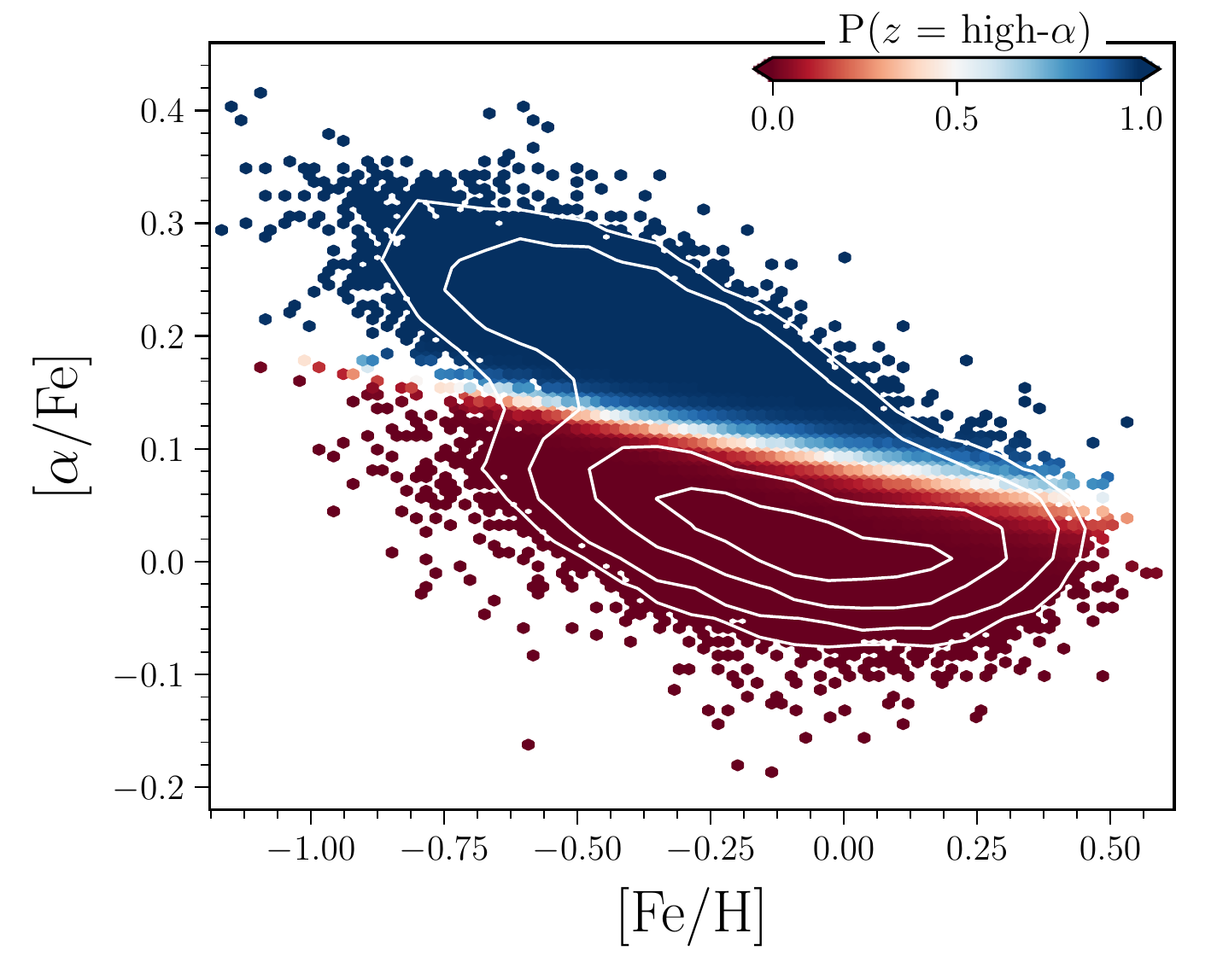}
\caption{Soft clustering of the \texttt{APOGEE} stars in the [$\alpha$/Fe]-[Fe/H] plane. Each bin is colored by the mean assignment ($z$) probability of the stars contained within the bin, where blue indicates a high probability of being assigned to the high-$\alpha$ sequence and red indicates a high probability of being assigned to the low-$\alpha$ sequence. The contours indicate the density of stars (with levels indicated at 30, 100, 500, and 1500 stars).}
\label{fig:separate}
\end{figure}

Fixing the number of clusters to two, we fit the data with a Gaussian mixture model. We find the results of the clustering to be sensitive to the initialization of the component means, so we compute a 2D kernel density estimate (KDE) of the data and use the highest-density locations in each of the two sequences to initialize the means. After convergence, for each star we record the component assignment ($z$) probability that the star belongs to the high-$\alpha$ sequence $P$($z$ = high-$\alpha$). As seen in Figure \ref{fig:separate}, stars with high-$\alpha$ abundances are assigned with high probability to the same mixture component, and stars with low-$\alpha$ abundances are assigned with high probability to the other mixture component. As expected, stars with intermediate-$\alpha$ abundances at a fixed [Fe/H] are assigned a lower probability of belonging to either the low or high-$\alpha$ sequence. 

For the results throughout this paper, we use the assignment probabilities to weight the stars when making comparisons between the low- and high-$\alpha$ sequence. Since 92\% of the stars in our sample are assigned with high probability to either component (i.e. $P$($z$ = high-$\alpha$.) = 0-5\% or 95-100\%), only a small fraction of stars with more ambiguous component assignments are given less weight. 

\subsection{Demonstration that [Mg/Si] does not \\
simply trace [$\alpha$/Fe]}
\label{sec:extra}

In Section \ref{sec:introduction} we described the physical motivation for looking at the ratio of magnesium to silicon. We now demonstrate, in a quantifiable way, that there is additional information encoded in the ratio of Mg to Si beyond its relation to a global $\alpha$ abundance. We do this by building a model to predict the [$\alpha$/Fe] abundances of stars using two sets of input features. In the first instance we use $\vec{\textbf{X}}$ = ([Fe/H], [Mg/Fe], [Si/Fe]), and in the second instance we use $\vec{\textbf{X}}$ = ([Fe/H], [Mg/Si]). The former set contains the individual Mg and Si abundances, and the latter contains just the ratio of the two. 

We split the \texttt{APOGEE} sample described in Section \ref{sec:main_sample} into a training set (70\%) and a hold-out set (30\%). The training set is used for model training and hyperparameter tuning. The hold-out set, which is data not seen during training or model selection, is used to evaluate the performance of the final model. To select the best model, with the training set we perform a grid search over model hyperparameters with a 10-fold cross-validation, and ultimately select the model that results in the best average r$^{2}$ score. The r$^{2}$ score is computed as the standard coefficient of determination, r$^{2}$ = 1 - $\frac{1}{N\sigma^{2}}\sum_{i}$($y_{\rm true, i}$ - $y_{\rm pred, i}$)$^{2}$, where $y_{\rm true}$ and $y_{\rm pred}$ are the true and model-predicted values of the dependent variable, $N$ is the number of observations, and $\sigma^{2}$ is the variance of $\vec{y_{\rm true}}$. An r$^{2}$ score closer to 1 indicates that the model predicts the variation in $y_{\rm true}$ well, whereas an r$^{2}$ score of 0 indicates that the model does not capture any of the variation.

\begin{figure}[tp]
\centering
\includegraphics[width=1.1\columnwidth]{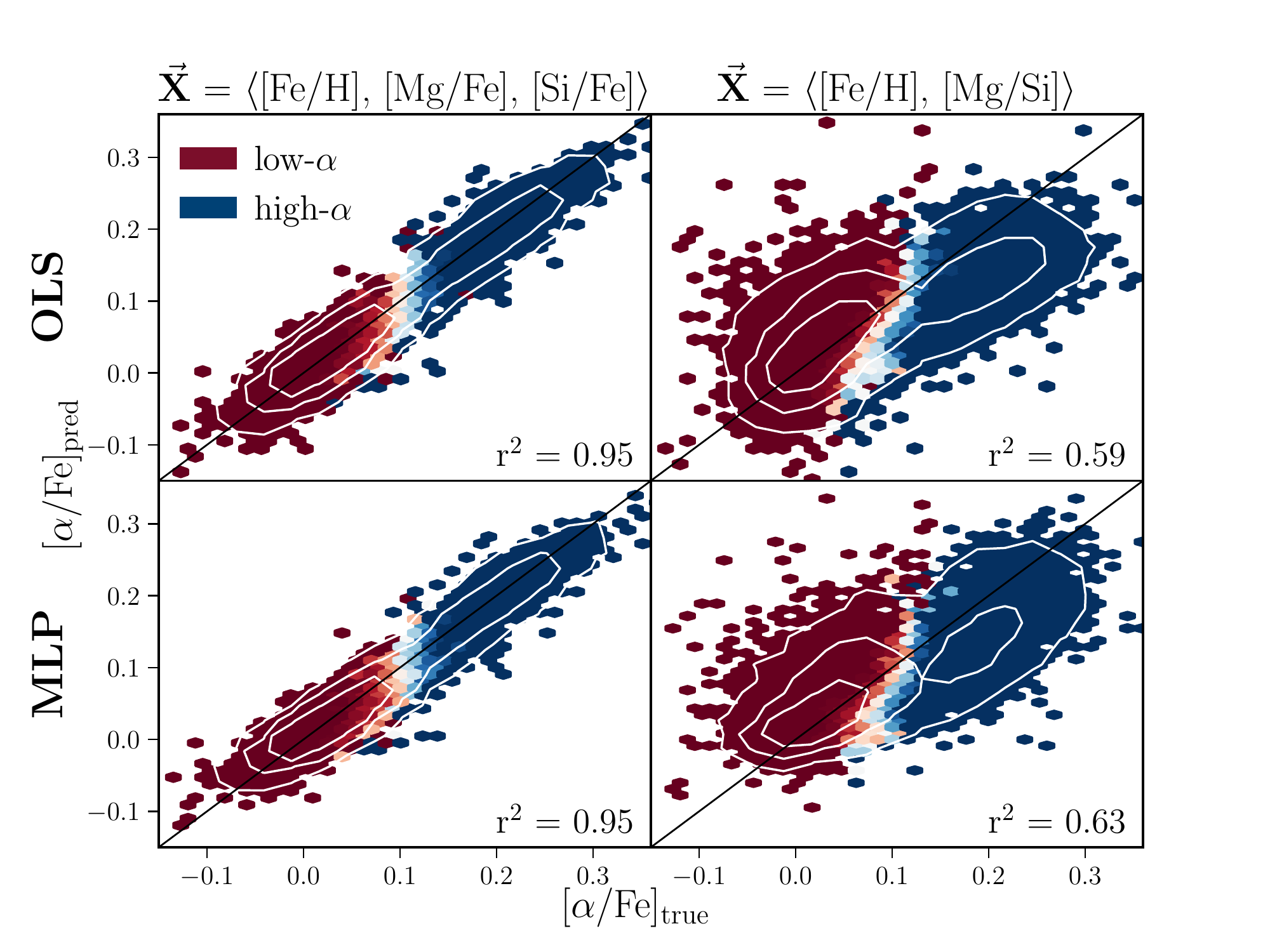}
\caption{Prediction of [$\alpha$/Fe] using two sets of input features, $\vec{\textbf{X}}$ = ([Fe/H], [Mg/Fe], [Si/Fe]) (left) and $\vec{\textbf{X}}$ = ([Fe/H], [Mg/Si]) (right). In each panel, the predicted vs. true [$\alpha$/Fe] abundances of stars in the hold-out set are shown, which are data not used during training and model selection. The binning color indicates the dominant $\alpha$ class, and the contours show the density of stars. The input vector $\vec{\textbf{X}}$ = ([Fe/H], [Mg/Si]) fails to predict the $\alpha$-enhancement of stars assuming both a linear model (OLS, top-row), and a more flexible model (MLP, bottom-row).}
\label{fig:pred_alpha}
\end{figure}

First we consider a linear model to predict [$\alpha$/Fe]. Since the dimensionality of the input feature space is low, we first implement unregularized ordinary least squares (OLS). The top row of Figure \ref{fig:pred_alpha} shows the results of the OLS model applied to the hold-out set. Using the individual Mg and Si abundances, we find that even this simple linear model predicts [$\alpha$/Fe] quite well, with an r$^{2}$ = 0.95. However, the results are considerably worse when using the input feature vector of $\vec{\textbf{X}}$ = ([Fe/H], [Mg/Si]). The r$^{2}$ is reduced to 0.59 and, as seen in the figure, the model tends to over-predict the $\alpha$ abundances of low-$\alpha$ sequence stars, and under-predict those high-$\alpha$ sequence stars. 

One possible reason why the [Mg/Si] abundance does not predict [$\alpha$/Fe] well is that the relationship is not captured by a linear model. To test this, we also train a vanilla feed-forward neural network (also referred to as a multilayer perceptron (MLP)) using both sets of input features. We create a sequential MLP model and perform a small grid search over several hyperparameters including: the hidden layer size (5, 10, 25, 100), the number of hidden layers (1, 2), the activation function (ReLU, tanh), and the regularization $\alpha$ (5 values from 10$^{-3}$ to 10$^{3}$). As seen in the bottom row of Figure \ref{fig:pred_alpha}, we find that an MLP model only performs marginally better (r$^{2}$ = 0.63) than the OLS model at predicting the [$\alpha$/Fe] abundances of the hold-out set from $\vec{\textbf{X}}$ = ([Fe/H], [Mg/Si]).

\begin{figure*}[tp]
\centering
\includegraphics[width=2.1\columnwidth]{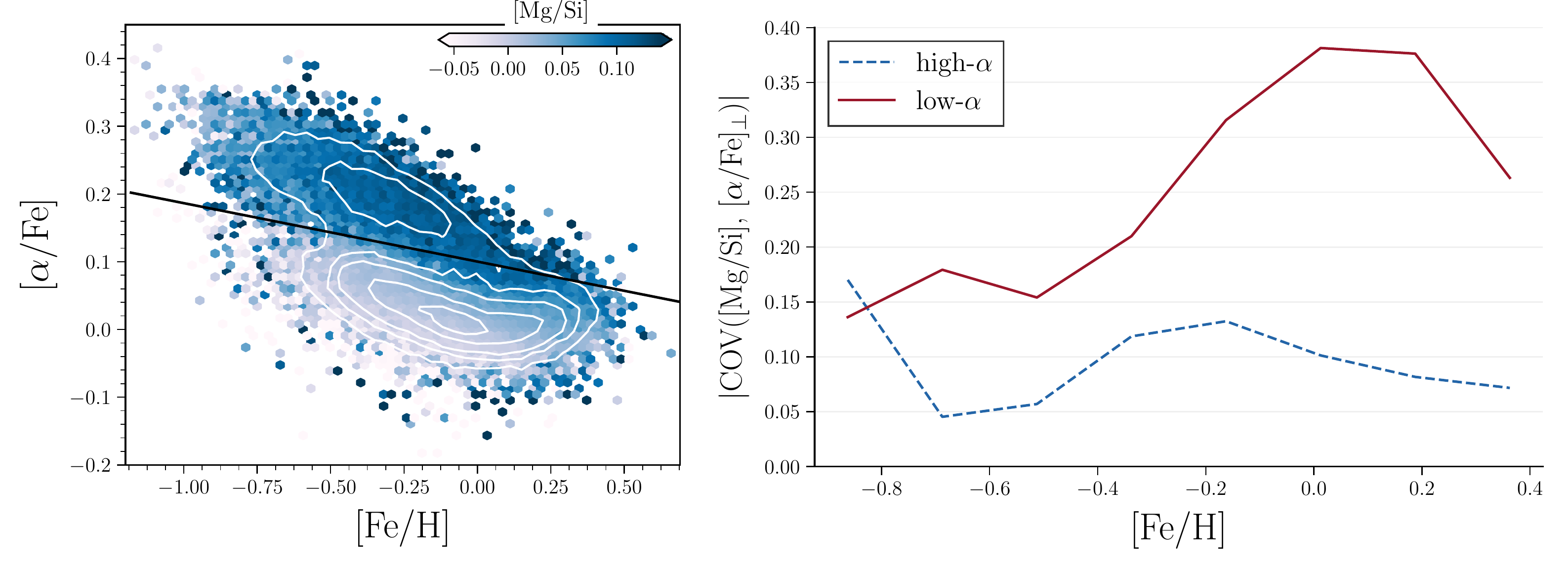}
\caption{\textbf{\textit{Left:}} [$\alpha$/Fe]-[Fe/H] distribution for the $\sim$70,000 stars in our \texttt{APOGEE} subsample. The bins are colored by the [Mg/Si] abundance, while the contours represent the density of stars. Here, we see that the high-$\alpha$ sequence stars have higher [Mg/Si] abundances than low-$\alpha$ sequence stars. \textbf{\textit{Right:}} absolute value of the covariance between [Mg/Si] and [$\alpha$/Fe]$_{\perp}$ as a function of metallicity for the low-$\alpha$ (red) and high-$\alpha$ (blue) sequences. In the low-$\alpha$ sequence, [Mg/Si] and [$\alpha$/Fe]$_{\perp}$ jointly vary more strongly than in the high-$\alpha$ sequence.}
\label{fig:mgsi_ratio}
\end{figure*}

To test if the dimensionality of the input feature vector is driving the difference in performance between the models trained with the two sets of features described above, we trained models with three additional sets of input features: (1) $\vec{\textbf{X}}$ = ([Fe/H], [Mg/Fe]), (2) $\vec{\textbf{X}}$ = ([Fe/H], [Si/Fe]), and (3) $\vec{\textbf{X}}$ = ([Fe/H], average([Mg/Fe], [Si/Fe])). Training with (1) we find the OLS r$^{2}$ to be 0.93 and the MLP r$^{2}$ to be 0.93. Training with (2) we find the OLS r$^{2}$ to be 0.74 and the MLP r$^{2}$ to be 0.77. And training with (3) we find the OLS r$^{2}$ to be 0.94 and the MLP r$^{2}$ to be 0.94. From this experiment we find that for this specific problem, the difference in the dimensionality of the input feature vector only marginally effects the prediction performance because [Mg/Fe] more strongly traces [$\alpha$/Fe] than [Si/Fe].

In the end, using both a simple linear model and a more flexible nonlinear model, the individual [Mg/Fe] and [Si/Fe] abundances predict the $\alpha$-enhancements of stars very well. This is expected, because the global $\alpha$ abundance is computed from information about the individual $\alpha$-element abundances. However, the predictive ability of the ratio of two $\alpha$-elements, [Mg/Si], is significantly worse. Even with a more flexible nonlinear model, [Mg/Si] does not predict the global $\alpha$-enhancement of stars as well as the individual [Mg/Fe] and [Si/Fe] abundances do. What this suggests is that there is additional information contained in the residuals of the prediction when using [Mg/Si] that is distinct from the two elements' relationship to a bulk $\alpha$ abundance. However, this demonstration does not tell us what the residuals from the [Mg/Si] prediction do trace. In this paper we investigate this information and seek to understand what [Mg/Si] reveals about the chemical evolution of the Milky Way.

\section{Empirical characterization of [Mg/Si]}
\label{sec:results}

\subsection{The [Mg/Si] abundance of the low- and high-$\alpha$ sequences}
\label{sec:first_ratio}
We begin our investigation by first characterizing the ratio of Mg to Si for both the low- and high-$\alpha$ sequences. The \textit{left panel} of Figure \ref{fig:mgsi_ratio} shows the [$\alpha$/Fe]-[Fe/H] distribution for the sample of $\sim$70,000 \texttt{APOGEE} stars described in Section \ref{sec:data}, where the bins are colored by the mean [Mg/Si] of the stars that fall within each bin. For the entire sample, the inner 90$^{\mathrm{th}}$ percentile of the [Mg/Si] abundances spans nearly 0.2 dex from [Mg/Si]= -0.05 - 0.14 dex, indicating that the stars have varying [Mg/Fe] and [Si/Fe] abundances. The median [Mg/Fe] and [Si/Fe] values for the high-$\alpha$ sequence are 0.25 dex and 0.1 dex, respectively. The median [Mg/Fe] and [Si/Fe] values for the low-$\alpha$ sequence are 0.035 dex and 0.015 dex, respectively. These differences in Mg and Si abundances lead to a mean [Mg/Si] $\sim$ 0.096 dex for the high-$\alpha$ sequence and mean [Mg/Si] $\sim$ 0.02 dex for the low-$\alpha$ sequence.

To quantify how the ratio of Mg to Si varies for stars \textit{within} the low- and high-$\alpha$ sequences, the \textit{right panel} of Figure \ref{fig:mgsi_ratio} shows how [Mg/Si] varies with [$\alpha$/Fe] across the [$\alpha$/Fe]-[Fe/H] plane. Here, we compute the absolute value of the covariance between [Mg/Si] and [$\alpha$/Fe]$_{\perp}$ in eight [Fe/H] bins from -1.0 $\le$ [Fe/H] $\le$ +0.5 dex, where [$\alpha$/Fe]$_{\perp}$ is measured perpendicular to a linear class boundary separating the low- and high-$\alpha$ sequences. In computing the covariance for the high- and low-$\alpha$ sequence separately, we weight the stars according to their cluster assignment probabilities as described in Section \ref{sec:clustering}. As seen in Figure \ref{fig:mgsi_ratio}, [Mg/Si] and [$\alpha$/Fe] jointly vary more strongly for low-$\alpha$ sequence stars than for high-$\alpha$ sequence stars across nearly the entire range of metallicities. The absolute value of the covariance for the high-$\alpha$ sequence never surpasses $\sim$0.2, whereas for the low-$\alpha$ sequence it reaches $\sim$0.4 at [Fe/H] $\sim$ 0.2 dex. This shows that the ratio [Mg/Si] behaves differently in the low- and high-$\alpha$ sequences, which suggests that this ratio could be probing differences in the chemical enrichment histories of the two sequences. 

Before continuing, we corroborate the finding that the high-$\alpha$ sequence has an excess of Mg relative to Si compared to the low-$\alpha$ sequence, with both the \texttt{APOGEE} RC sample described in Section \ref{sec:rc} and the HARPS sample described in Section \ref{sec:harps}. First, to obtain cluster membership probabilities for both of these datasets, we apply the same Gaussian mixture model trained on the full \texttt{APOGEE} sample. For the RC stars, the high-$\alpha$ sequence is found to have a mean [Mg/Si] of 0.07 dex, while the low-$\alpha$ sequence has a mean [Mg/Si] of 0.003 dex. For the HARPS stars, the mean [Mg/Si] of the high-$\alpha$ sequence is 0.07 dex and the mean [Mg/Si] of the low-$\alpha$ sequence is 0.01 dex. These differences are similar to what is found for the full \texttt{APOGEE} sample. After verifying that the [Mg/Si] trends are robust to log $g$ and $T_{\mathrm{eff}}$ variations, and are also present in a sample of stars with higher-quality spectra, in the following section we explore how [Mg/Si] evolves with stellar age and metallicity.

\subsection{[Mg/Si] trends with age and metallicity}
\label{sec:age_metallicity}
To further examine differences in the [Mg/Si] abundance between the low- and high-$\alpha$ sequences, we now consider variations with other stellar properties available to us, including ages and [Fe/H]. Investigating correlations with these additional parameters will allow us to build intuition for how the production of Mg relative to Si varies both through time and in different star-formation environments. 

As described in Section \ref{sec:main_sample}, we use stellar age estimates from \cite{ness16}, which are derived from CN absorption lines and are accurate to $\sim$40\%. Considering these uncertainties, we divide our sample of \texttt{APOGEE} stars into three wide-age bins, roughly separating young stars born 0 - 3 Gyr ago, intermediate-aged stars born 3 - 7 Gyr ago, and old stars born 7 - 14 Gyr ago. This binning results in 561 young stars, 3,981 intermediate-aged stars, and 8,254 old stars in the high-$\alpha$ sequence (P($z$ = high-$\alpha$) \textgreater~0.95), and 19,957 young stars, 23,685 intermediate-aged stars, and 8,747 old stars in the low-$\alpha$ sequence (P($z$ = high-$\alpha$) \textless~0.05). The median age of stars in the high-$\alpha$ sequence is 8.26 Gyr with a standard deviation of 3.0 Gyr, while the median age of stars in the low-$\alpha$ sequence is 3.7 Gyr with a standard deviation of 2.8 Gyr. There is only a relatively small fraction of high-$\alpha$ stars that are young, and old low-$\alpha$ stars that are old (e.g. see \cite{silva18}). These have corresponding abundance and age errors that indicate these measurements are robust. As a consequence of our large sample, we have a sufficient numbers of stars to examine the trends of the high- and low-$\alpha$ sequence to high significance, across all ages.

\begin{figure}[tp]
\includegraphics[width=1\columnwidth]{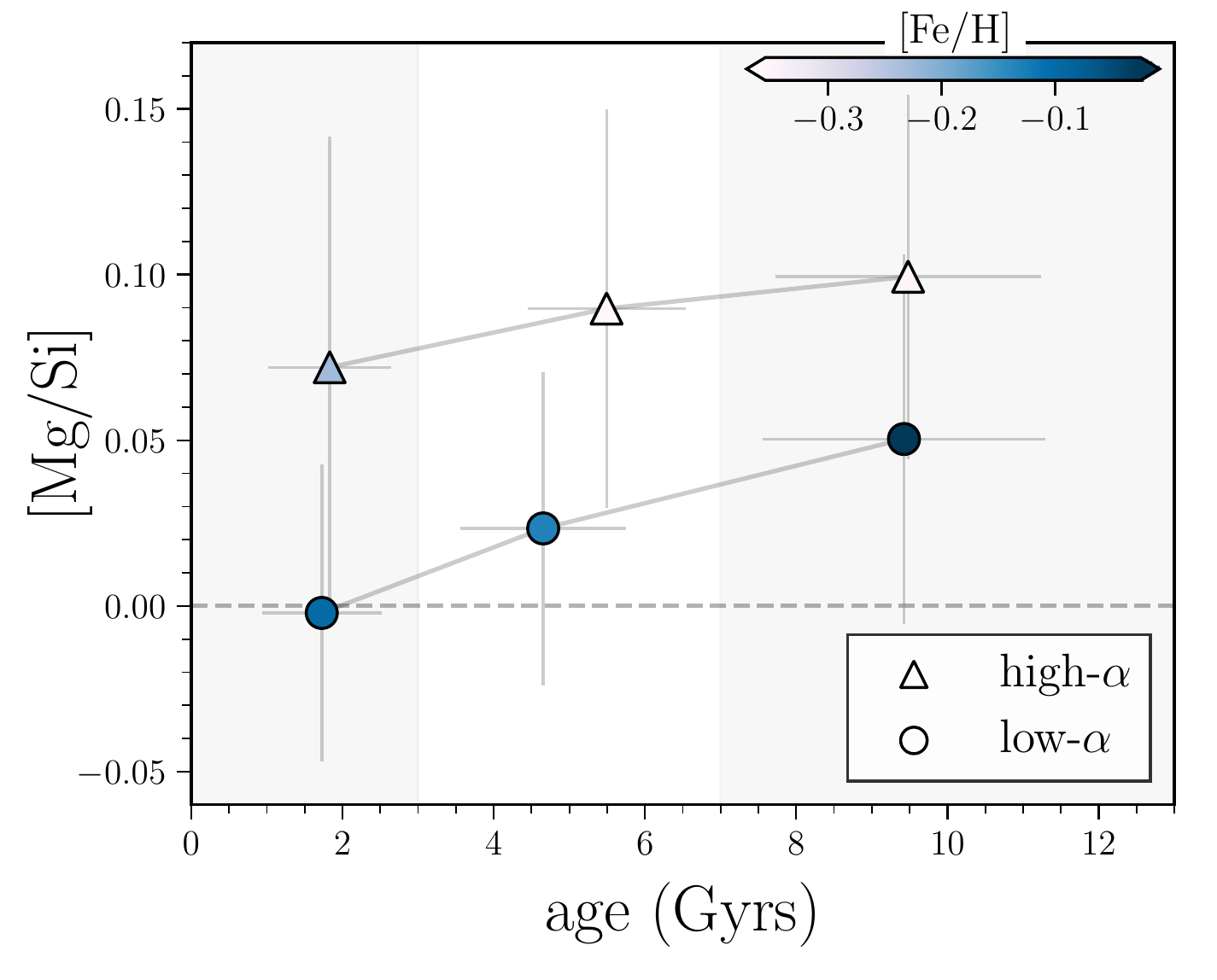}
\caption{The [Mg/Si] abundance of the low-$\alpha$ (circles) and high-$\alpha$ (triangles) sequences as a function of stellar age. The gray error bars indicate the  standard deviations of [Mg/Si] and age in each stellar age bin, while the errors on the mean are smaller than the size of the symbols.}
\label{fig:allstars_age}
\end{figure}

Figure \ref{fig:allstars_age} shows the mean [Mg/Si] abundance in each age bin, for both the low- and high-$\alpha$ sequences. As expected from Section \ref{sec:first_ratio}, in each bin the high-$\alpha$ sequence has a higher mean [Mg/Si] than the low-$\alpha$ sequence. It should be noted that while the differences in the mean [Mg/Si] values are significant, specifically the standard errors on the means are smaller than the size of the symbols in the figure, the 1$\sigma$ dispersions around the mean of the distributions (indicated in gray) do overlap. So, while the distributions peak at different values, they are not completely disparate. The takeaway of Figure \ref{fig:allstars_age} is the trends we observe with age and the mean [Mg/Si]. Without yet taking metallicity into consideration, we notice two things. The first is that stars born earlier in time (older ages) are more enhanced in Mg relative to Si than stars born at later times (younger ages). This is consistent with the relative theoretical Mg/Si yields discussed in Section \ref{sec:theory}. The second observation is the mean [Mg/Si] abundance varies more with age for the low-$\alpha$ sequence stars than for the high-$\alpha$ sequence stars. The low-$\alpha$ sequence spans $\sim$0.049 dex in [Mg/Si] from the young age bin to the old age bin, while the high-$\alpha$ sequence only spans $\sim$0.026 dex. We also verify that this trend with age is present in the RC sample. Considering the same age bins, the [Mg/Si] of low-$\alpha$ sequence stars decreases by 0.038 dex from the oldest to youngest stars. For the high-$\alpha$ sequence stars, this decrease in [Mg/Si] with age is only 0.023 dex. We cannot confirm this [Mg/Si] age trend with the HARPS data because our sample only includes eight $\alpha$-enriched stars, all with ages between 8.1 and 9 Gyr.

\begin{figure*}[tp!]
\centering
\includegraphics[width=2.1\columnwidth]{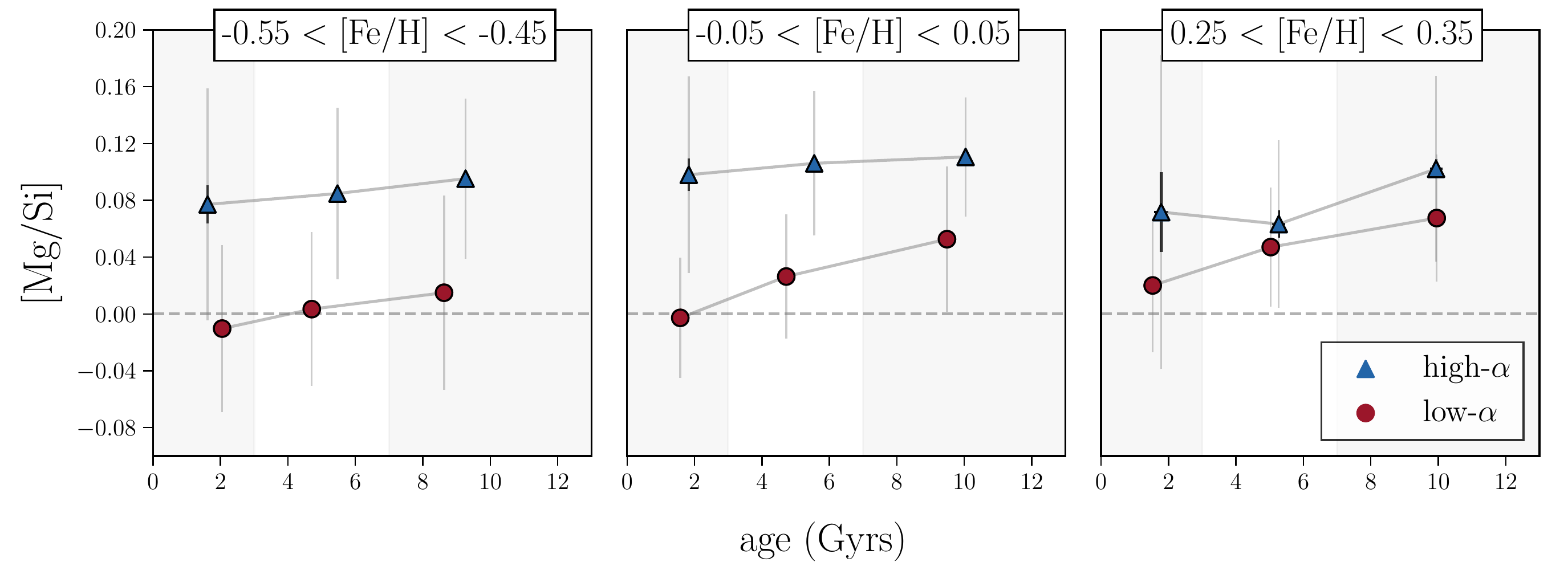}
\caption{[Mg/Si] abundance as a function of stellar age in three narrow metallicity bins of 0.1 dex. The left panel shows a metal-poor bin, the middle panel shows a solar-metallicity bin, and the right panel shows a metal-rich bin. In each panel, the high-$\alpha$ sequence is shown in blue, and the low-$\alpha$ is shown in red. The light gray error bars represent the 1$\sigma$ standard deviation of [Mg/Si] in each age bin, while the black error bars represent the error on the mean.}
\label{fig:met_bins}
\end{figure*}

Figure \ref{fig:allstars_age} also reveals that metallicity varies with age and [Mg/Si]. For high-$\alpha$ sequence stars, younger stars are more metal-rich than older stars, with the average [Fe/H] decreasing by $\sim$0.12 dex with increasing stellar age. However, for low-$\alpha$ sequence stars, younger stars are instead more metal-poor than older stars, with the average [Fe/H] increasing by $\sim$0.07 dex with increasing stellar age. The opposite [Fe/H] age trend between the low- and high-$\alpha$ sequences is a further line of evidence that the chemical enrichment histories of the two sequences is different. To understand the variation of [Mg/Si] with age independently of metallicity, we condition on [Fe/H] by examining how the [Mg/Si] age trend varies is narrow metallicity bins. Figure \ref{fig:met_bins} shows three 0.1 dex wide [Fe/H] bins: a metal-poor bin (-0.55 \textless ~[Fe/H] \textless~-0.45 dex), a solar metallicity bin (-0.05 \textless ~[Fe/H] \textless~0.05 dex), and a metal-rich bin (0.25 \textless ~[Fe/H] \textless~0.35 dex). 

In each metallicity bin, we see that the [Mg/Si] abundance of the low-$\alpha$ sequence increases with stellar age. This suggests that the general trend of the relationship is independent of variations in [Fe/H]. However, the normalization of the trend does vary with metallicity. Stars in the metal-rich bin have overall higher [Mg/Si] abundances than stars in the metal-poor bin. In contrast to the low-$\alpha$ sequence, the [Mg/Si] abundance of the high-$\alpha$ sequence remains relatively constant with age in the [Fe/H] bins. This is the case in the metal-poor and solar-metallicity bin, where similar [Mg/Si] values are found at each age. However, in the metal-rich bin there is some variation with the oldest high-$\alpha$ sequence stars having larger ratios of Mg to Si than the intermediate-age and young high-$\alpha$ sequence stars. Part of this trend can be attributed to the lack of separation between the two sequences at higher metallicities. As seen in Figure \ref{fig:separate}, at metallicies higher than [Fe/H] $\sim$ 0.0 dex, the high-$\alpha$ sequence appears to merge into the low-$\alpha$ sequence. This makes the definition of the two sequences more ambiguous at these high metallicities, which is reflected in assignment probabilities discussed in Section \ref{sec:clustering}. 

Based on Figure \ref{fig:met_bins}, we can begin to hypothesize about the origin of these [Mg/Si] abundance trends by considering the nucleosynthetic channels of the two elements. As discussed in Section \ref{sec:introduction}, CC-SN produce both Mg and Si. However, Si is also produced in SN Ia. Given this, one explanation for the decrease in [Mg/Si] with age for the low-$\alpha$ sequence is that this is an imprint of the time-dependent yield contributions from SN Ia, which we expect from Figure \ref{fig:theory_mgsi}. Since CC-SN enrichment occurs instantaneously compared to SN Ia enrichment, the Mg abundance is relatively constant in time. Therefore, for the low-$\alpha$ sequence a possible reason for the decrease in [Mg/Si] with age is the steady increase in Si over time. What this could imply about the high-$\alpha$ sequence, where [Mg/Si] is constant with stellar age, is that the environment in which these stars form is unpolluted by SN Ia ejecta at all times. While we expect the oldest (and majority) of the high-$\alpha$ sequence stars to be unaffected by SN Ia, what Figure \ref{fig:met_bins} suggests is that even the younger high-$\alpha$ sequence stars are unpolluted by Si from SN Ia. This could indicate that at all star-formation epochs, the formation of low- and high-$\alpha$ sequence stars occurs distinctly. Nonetheless, there could be numerous alternative explanations for the trends we find in Figure \ref{fig:met_bins}. One is that the Mg enrichment of the gas from which high-$\alpha$ sequence stars are formed could steadily increase over time to match the increasing Si enrichment. This would result in a flat [Mg/Si] age trend without requiring isolation from SN Ia pollution. A possible mechanism for this increase in Mg enrichment is if high-$\alpha$ sequence stars formed according to an IMF with a high-mass end slope that became flatter over time. This would produce relatively more high-mass stars which, as discussed in Section \ref{sec:introduction}, would yield more Mg than Si. We attempt to explore these possibilities in Section \ref{sec:chempy} through GCE modeling.

\subsection{Spatial and orbital trends with [Mg/Si]}
\label{sec:spatial_orbits}

\subsubsection{Trends with Galactic location}
\label{sec:spatial}
We now turn toward empirically characterizing the relationship between [Mg/Si] and disk structure and dynamics. We do this by establishing how the Mg to Si ratio varies with location throughout the disk, and by investigating how the stellar actions are related to [Mg/Si]. By making this connection between the chemistry and the structural and orbital properties of stars, we hope to understand how [Mg/Si] might encode unique information regarding the formation and build-up of the Milky Way disk. 

To start, we consider how [Mg/Si] varies with Galactocentric radius (R) and distance from the disk midplane ($|z|$). As discussed in Section \ref{sec:main_sample}, we match our sample of \texttt{APOGEE} stars with the \cite{sanders} catalog to obtain the coordinates of each star. Inspired by Figure 4 of \cite{hayden15}, we divide the sample into three $|z|$ bins: 0 - 0.5 kpc, 0.5 - 1.0 kpc, and 1.0 - 2.0 kpc. Each of these three bins in $|z|$ is further divided into six radius bins, ranging from 3 to 15 kpc in 2 kpc wide annuli. To quantify the impact of each star's distance measurement uncertainty on the ($|z|$, R) binning, we perform a Monte Carlo simulation. In short, during each Monte Carlo iteration every star's coordinates are re-sampled from

\begin{equation}
z_{i} \sim \mathcal{N}(|z|, z_{\rm err})
\end{equation}

\begin{equation}
R_{i} \sim \mathcal{N}(\mathrm{R}, \mathrm{R}_{\rm err})
\medskip
\end{equation}

\noindent where the coordinates are described as normal distributions centered at the \cite{sanders} derived values, with a variance equal to the reported errors. After sampling $z_{i}$ and $R_{i}$ for each star, the stars are re-binned by ($|z|$, R) and the mean [Mg/Si] and [Fe/H] values of each bin are recorded. Repeating this procedure N = 10$^{3}$ times, we obtain the median and range of the mean [Mg/Si] and [Fe/H] values in each bin. 

\begin{figure*}[tp!]
\centering
\includegraphics[width=1.8\columnwidth]{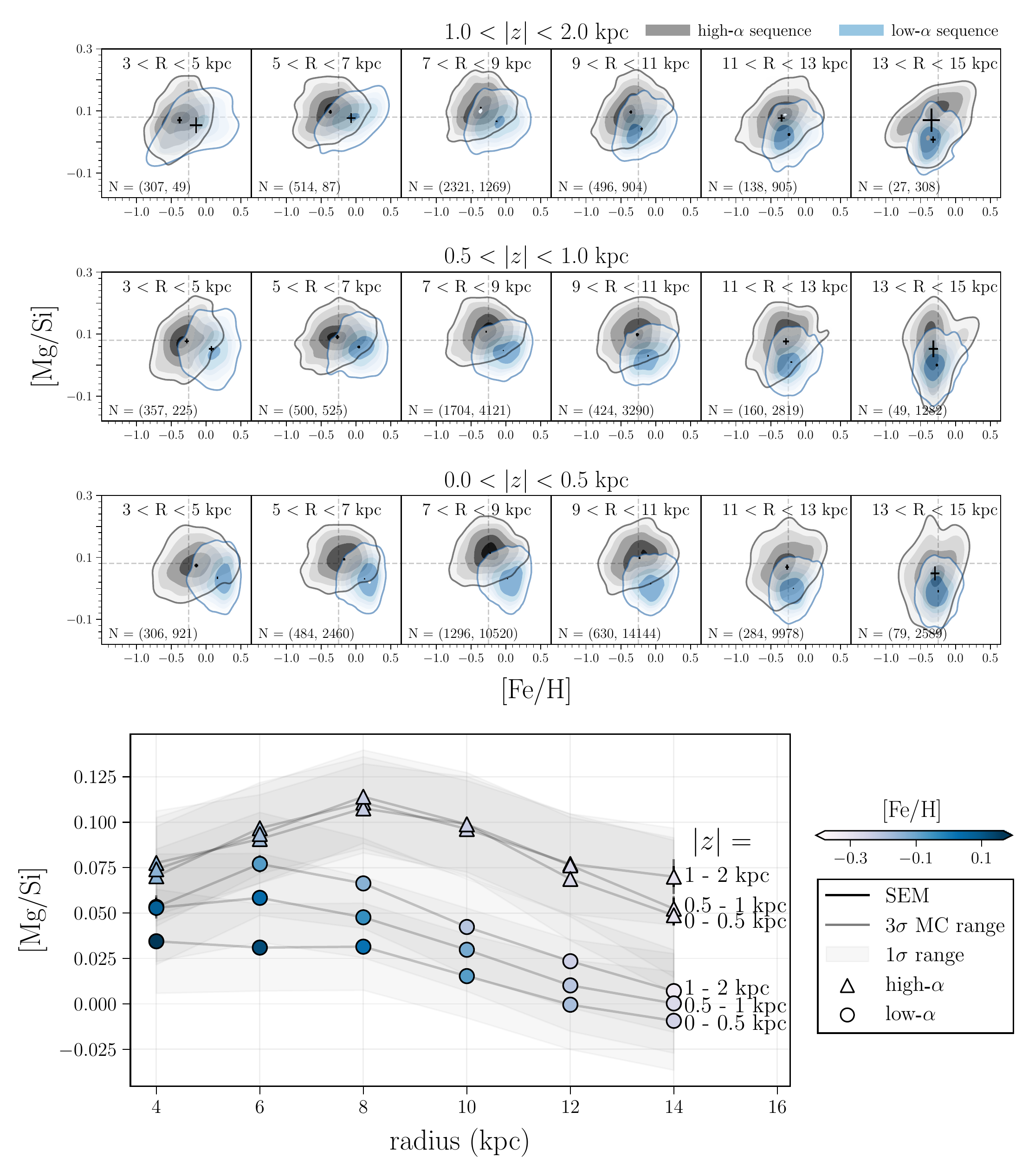}
\caption{\textbf{\textit{Top:}} distribution of high-$\alpha$ (gray) and low-$\alpha$ (blue) sequence stars in the [Mg/Si]-[Fe/H] plane as a function of Galactocentric radius (R) and distance from the disk midplane ($|z|$). Each row shows the distribution in 2 kpc wide radius bins, from the inner disk (left) to the outer disk (right). The bottom row shows the stars closest to the disk midplane from $|z| = 0 - 0.5$ kpc, the middle row shows stars from $|z| = 0.5 - 1$ kpc, and the top row shows stars farthest from the midplane at $|z| = 1 - 2$ kpc. In each panel the error bars represent the range of the mean [Mg/Si]-[Fe/H] values based on 10$^{3}$ Monte Carlo samplings of R and $|z|$. The number of stars in each bin is indicated as N = (\# high-$\alpha$ stars, \# low-$\alpha$ stars). \textbf{\textit{Bottom:}} this figure summarizes the trends shown in the panel above, showing the mean [Mg/Si] as a function of radius for each R,$|z|$ bin. The black error bars represent the standard error on the mean, the gray error bars represent the 3$\sigma$ range from the Monte Carlo simulation, and the gray bands represent the 1$\sigma$ range of the [Mg/Si] distribution. Here, trends with [Mg/Si], [Fe/H], and position within the Galaxy are more apparent. For high-$\alpha$ sequence stars (triangles) the peak [Mg/Si] occurs at an intermediate disk radius of $\sim$ 8 kpc, regardless of height from the disk midplane. For low-$\alpha$ sequence stars (circles), the peak [Mg/Si] occurs more toward the inner disk at $\textless$ 6 kpc, and in each radius bin stars at larger distances from the disk midplane are more enhanced in [Mg/Si] than stars residing closer to the disk midplane.}
\label{fig:disk}
\end{figure*}

The \textit{top panel} of Figure \ref{fig:disk} shows the low- and high-$\alpha$ sequence [Mg/Si]-[Fe/H] distributions in the different ($|z|$, R) bins, as well as the results of the Monte Carlo sampling simulation described above. In the Figure, the displayed contours are based on the mean posterior distances to each star, while the error bars represent the results of the Monte Carlo sampling. First we examine the high-$\alpha$ sequence. At all distances from the midplane, the [Mg/Si]-[Fe/H] distribution evolves similarly from the inner disk to the outer disk, where the peak of the distribution appears to increase marginally in [Mg/Si] abundance from the inner disk (3 kpc) to the mid-disk (7 - 9 kpc), and then decreases toward the outer disk (15 kpc). However, in the outermost region of the disk considered (13 - 15 kpc) the distance errors combined with the limited sample size significantly impact the binning. 

In contrast to the high-$\alpha$ sequence, the low-$\alpha$ sequence displays more significant spatial [Mg/Si]-[Fe/H] trends. As seen in bottom row of the \textit{top panel} of Figure \ref{fig:disk}, close to the disk midplane ($|z|$ = 0 - 0.5 kpc) the peak of the [Mg/Si] distribution decreases from $\sim$0.02 dex in the inner part of the Galaxy to $\sim$-0.01 dex in the outer regions. While this gradient in [Mg/Si] with radius is weak, it is significant compared to the stellar distance uncertainties and sample sizes of each bin. A similar trend with radius is seen farther from the midplane. This figure shows that the low-$\alpha$ sequence stars currently residing in the inner region were formed from gas that was more enriched in Mg relative to Si compared to the stars currently residing in the outer disk. 

The \textit{bottom panel} of Figure \ref{fig:disk} summarizes the mean trends seen in the \textit{top panel} of Figure \ref{fig:disk}, for both the low- and high-$\alpha$ sequences. The [Mg/Si]-radius relationship is shown for each bin in $|z|$ and the marker color indicates the mean [Fe/H]. The shaded gray regions indicate the 1$\sigma$ distribution, the black error bars indicate the standard error on the mean, and the gray error bars indicate the 3$\sigma$ range from the Monte Carlo simulation described above. While the information shown in this figure is the same as that in the \textit{top panel} of Figure \ref{fig:disk}, this representation of the data better lends itself to visualizing the relevant trends. First consider the high-$\alpha$ sequence. For these stars we see that the mean [Fe/H] decreases from the inner to the outer disk, and that the trend with mean [Mg/Si] and radius is similar regardless of distance from the disk midplane. Here, we see more clearly that high-$\alpha$ sequence stars in the intermediate disk (at $\sim$8 kpc) have the highest [Mg/Si] abundances on average ($\sim$ 0.11 dex) compared to the average [Mg/Si] abundances of stars in the inner disk ($\sim$0.07 dex) and outer disk ($\sim$0.06 dex). This trend of an average [Mg/Si] which first increases from $\sim$3 kpc to $\sim$8 kpc, and then decreases from $\sim$8 kpc to $\sim$15 kpc, is robust to the standard errors on the mean (which span lengths smaller than the height of the markers in the figure). 

The low-$\alpha$ sequence stars exhibit markedly different behavior with mean [Mg/Si] and ($|z|$, R). As seen in the \textit{bottom panel} of Figure \ref{fig:disk}, while the shape of the [Mg/Si]-radius curve is similar for all bins in $|z|$, at each radius stars that reside farther from the disk midplane have higher [Mg/Si] abundances compared to those that reside closer to the disk midplane. For instance, in the R = 5 - 7 kpc bin, low-$\alpha$ stars residing at $|z|$ = 0 - 0.05 kpc have an average [Mg/Si] abundance of 0.03 dex, while the stars residing at $|z|$ = 1 - 2 kpc have an average [Mg/Si] abundance of 0.076 dex. Compared to the high-$\alpha$ sequence stars, the shape of the mean [Mg/Si]-radius trends are also different. For $|z|$ \textgreater~0.5 kpc, the stars with the highest [Mg/Si] abundances on average are located at R $\sim$ 6 kpc, and then the mean [Mg/Si] decreases with radius until R $\sim$ 15 kpc. For the stars closest to the disk midplane ($|z|$ \textless~0.5 kpc), the mean [Mg/Si] is nearly constant from R $\sim$ 3 - 9 kpc, and then decreases until R $\sim$ 15 kpc. To test the robustness of these observed trends to variations in stellar properties, namely in $T_{\rm eff}$, we examined the [Mg/Si]-radius relationship considering stars with temperatures constrained to 200 K wide bins across 4300 K to 5100 K. We find that the general trends presented in the bottom panel of Figure \ref{fig:disk} hold across these narrower temperature ranges.

The main takeaways from this exploration of how [Mg/Si] varies throughout the disk are summarized as follows. First, from the \textit{top panel} of Figure \ref{fig:disk} we learn that, similar to how the distribution of [$\alpha$/Fe]-[Fe/H] varies with location in the disk (as in Figure 4 of \cite{hayden15}), the distribution of [Mg/Si]-[Fe/H] also varies with Galactic coordinates. We examine the high- and low-$\alpha$ sequence stars separately, and observe that, while their [Mg/Si]-[Fe/H] distributions significantly overlap, there are small differences between how the distributions of the two sequences change with R and $|z|$. From the \textit{bottom panel} of Figure \ref{fig:disk} we learn that the large number of stars in each bin affirms that the differences between the two $\alpha$ sequences in their mean [Mg/Si]-spatial trends are significant. 

The trends seen with mean [Mg/Si] abundance and $|z|$ for the low-$\alpha$ sequence stars could suggest that the low-$\alpha$ sequence part of the disk was built both up and out from layers of gas that had varying overall [Mg/Si] normalizations that scale with $|z|$, but similar gradients with respect to radius. On the other hand, the high-$\alpha$ sequence trends suggest that the gas that formed these stars was similarly enriched in Mg and Si at all distances from the disk midplane. These observations are consistent with how the ages of stars vary throughout the disk for both the low- and high-$\alpha$ sequences. Considering the same spatial bins defined in Figure \ref{fig:disk}, at every $|z|$, low-$\alpha$ sequence stars that reside closer to the Galactic center are older than their counterparts at larger radii. And at fixed R, low-$\alpha$ sequence stars that reside closer to the disk midplane are younger than the stars that at greater heights from the midplane. However, for the high-$\alpha$ sequence stars, we find little to no trend with stellar age and Galactic coordinates. 

Aside from their different trends with $|z|$, the mean [Mg/Si] of the high- and low-$\alpha$ sequences also peak at different radii, with the low-$\alpha$ sequence stars having higher [Mg/Si] abundances more toward the inner disk and the high-$\alpha$ sequence stars having higher [Mg/Si] abundances in the intermediate disk. These differences must be reflective of the time sequence of star-formation, and of radial migration effects. Low-$\alpha$ sequence stars largely form in the disk midplane where there are gradients in [Mg/Si] and [Fe/H]. As time goes on, these stars form at progressively larger radii. On the other hand, high-$\alpha$ sequence stars form more toward the central regions of the Galaxy with higher and clumpier efficiencies. Since high-$\alpha$ sequence stars are generally older, they presumably have had more time to migrate both vertically and radially. Distinct radial migration patterns, for example caused by different dynamical effects of the Galactic bar on the low- versus high-$\alpha$ sequence populations, could also potentially result in the trends we observe. These broad speculations should be revisited in the context of a more rigorous modeling approach.

\begin{figure*}[tp!]
\centering
\includegraphics[width=2\columnwidth]{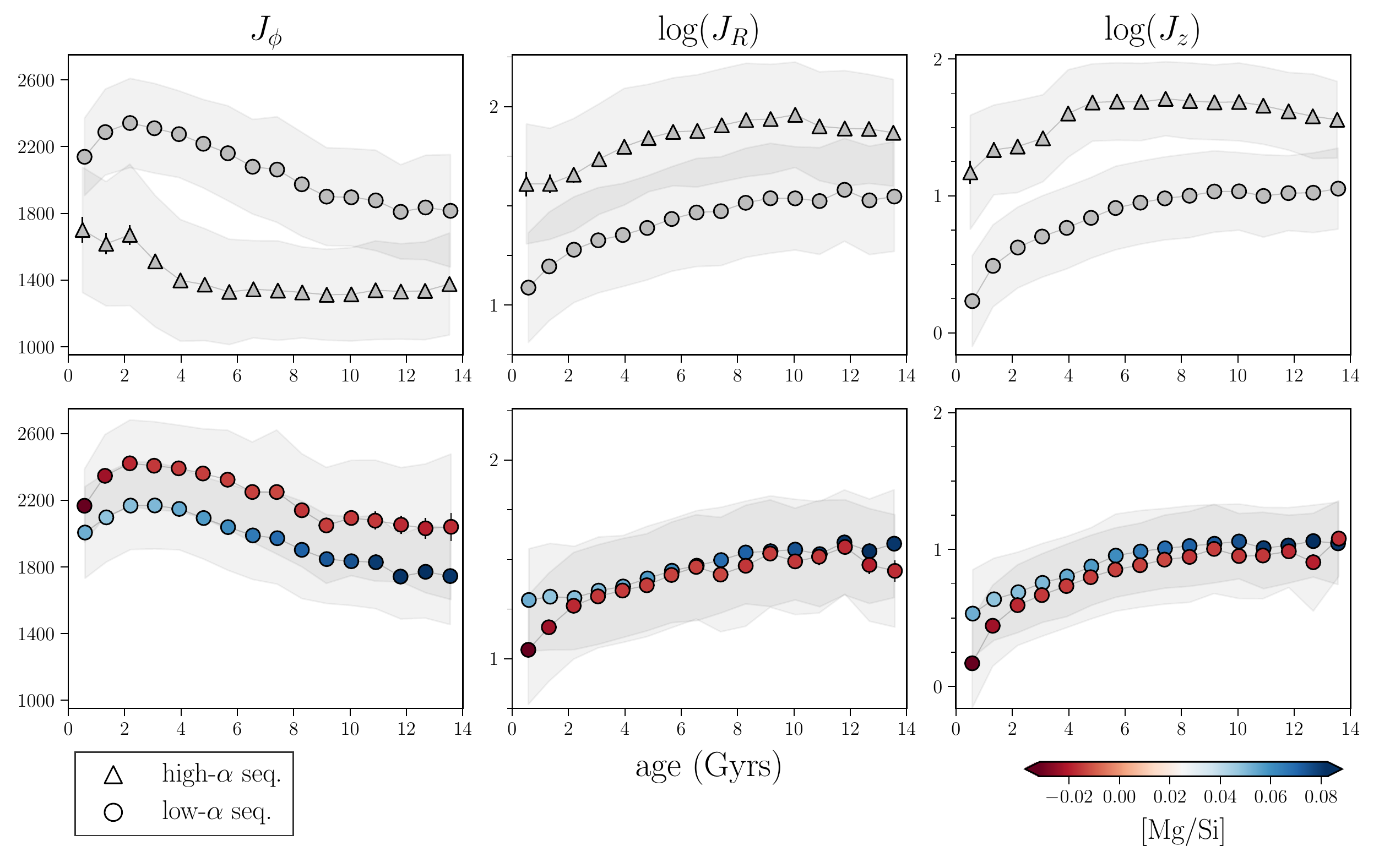}
\caption{\textbf{\textit{Top row}}: evolution of orbital actions ($J_{\phi}$, $J_{R}$, $J_{z}$ [kpc km s$^{-1}$]) with stellar age for low- and high-$\alpha$ sequence stars. For 16 age bins, the average action value is shown and in gray the 1$\sigma$ standard deviation is indicated. At every age, the actions of high- and low-$\alpha$ sequence stars are distinct. \textbf{\textit{Bottom row}}: low-$\alpha$ sequence action-age relations now divided into stars with lower vs. higher [Mg/Si] abundances. For $J_{\phi}$, part of the scatter in the relation with age is described by a gradient in [Mg/Si], where stars with lower [Mg/Si] values have higher angular momentum than those with higher [Mg/Si] values. }
\label{fig:actions}
\end{figure*}

\subsubsection{Trends with orbital properties}
\label{sec:orbits}
Aside from considering location within the Galaxy, we also characterize the relationship between [Mg/Si] and stellar orbital properties. We do this by examining how [Mg/Si] varies with the three actions, $J_{\phi}$, $J_{R}$, and $J_{z}$. As mentioned in Section \ref{sec:main_sample}, we obtain actions for our \texttt{APOGEE} sample from \cite{sanders}, who adopt a \cite{mcwilliam13} Milky Way potential to compute $J_{\phi}$, $J_{R}$, and $J_{z}$ from \textit{Gaia} astrometry using the St\"ackel Fudge method \citep{sanders16}. Assuming an axisymmetric and relatively time-independent potential, the three actions uniquely define a stellar orbit. Concise physical explanations of the three actions are given in \cite{trick}. The azimuthal action, $J_{\phi}$, quantifies a star's rotation about the center of the Galaxy. This quantity is the same as angular momentum in the vertical direction, and throughout this paper we refer to azimuthal action of a star as its angular momentum. The radial action, $J_{R}$, describes the amount of oscillation a star exhibits in the radial direction, which is related to the eccentricity of the orbit. Stars with $J_{R}$ = 0 are on circular orbits. Finally, the vertical action, $J_{z}$, measures the excursion of a star in the vertical direction, where $J_{z}$ = 0 indicates that a star is confined to the disk midplane. 

To begin, the top row of Figure \ref{fig:actions} shows how the actions of high- and low-$\alpha$ sequence stars differ across stellar ages. Similar to what \cite{gandhi} report for stars with \texttt{LAMOST} abundances, we find that the low- and high-$\alpha$ sequences are distinct in their actions at all ages. First considering angular momentum, we see that low-$\alpha$ sequence stars have higher $J_{\phi}$ values than high-$\alpha$ sequence stars at each age. This separation between the two sequences is robust to the standard errors on the mean $J_{\phi}$ in each age bin, and at most ages the distributions are separated by at least one standard deviation. The separation in $J_{\phi}$ between the two sequences is consistent with the notion that the low-$\alpha$ sequence comprises a more radially extended disk, while the high-$\alpha$ sequence is confined closer to the Galactic center. Since $J_{\phi}$ traces radial location, these trends are also reflective of a gas disk where star-formation proceeded outward over time with a radial gradient in chemical abundances. This is inside-out formation of the disk, where stars with lower $J_{\phi}$ values (residing in the inner disk) formed at early times, and since then the radial size of the disk was built up from progressively younger stars.

We also find the high- and low-$\alpha$ sequences to be distinct in radial action, $J_{R}$, and vertical action, $J_{z}$. As seen in the middle panel of the top row of Figure \ref{fig:actions}, in every age bin the high-$\alpha$ sequence stars exhibit larger $J_{R}$ values than the low-$\alpha$ sequence stars. This trend is robust to the standard errors on the mean; however, the distributions do overlap within 1$\sigma$. What these trends with $J_{R}$ suggest is that, at all ages, high-$\alpha$ sequence stars are on more eccentric orbits than low-$\alpha$ sequence stars, and that for both sequences older stars are described by more eccentric orbits than younger stars. One explanation for the relationships between radial action and age is that older stars have had more time for their orbits to be perturbed to more eccentric orbits, and this perturbation occurs regardless of a star's $\alpha$-enhancement. The observed trends with $J_{z}$ are similar to those with $J_{R}$. As seen in the figure, at every age the high-$\alpha$ sequence stars are described by larger vertical actions than the low-$\alpha$ sequence stars. This trend also evolves with age for both sequences, where older stars have higher $J_{z}$ values than younger stars. These trends are robust to the standard errors on the mean, and the distributions between the two sequences are distinct by at least 1$\sigma$. The differences in the $J_{z}$ values of the low- and high-$\alpha$ sequence stars is reflective of their general location within the Galaxy, where low-$\alpha$ sequence stars are more confined to the disk midplane and high-$\alpha$ sequence stars mostly comprise the ``thick", vertically extended disk of the Milky Way. 

We speculate why high-$\alpha$ sequence exhibit higher $J_{z}$ and $J_{R}$ values than low-$\alpha$ sequence stars at every age. One possibility is that because the low- and high-$\alpha$ sequences form in different spatial locations (and presumably from different reservoirs of gas), they subsequently experience distinct modes of star-formation that result in different initial orbital properties. An additional possibility is that heating processes may be more active in the thicker, inner disk where high-$\alpha$ sequence stars reside, which perturbs these stars to higher radial and vertical excursions. Lastly, dynamical times are shorter in the inner disk. So the dynamical ages of high-$\alpha$ stars are generally older than their stellar ages, which means they would have more time to be perturbed to higher $J_{z}$ and $J_{R}$ values.

Considering the evolution with age, as with the radial action, older stars have had more time to be perturbed to larger vertical excursions. However, while the $J_{z}$ values of low-$\alpha$ sequence stars increase steadily with stellar age, those of high-$\alpha$ sequence stars remain nearly constant from 4 - 14 Gyr. This suggests that $\alpha$-enhancement could be a signature of initial orbital properties, as well as the dynamical processes that perturb orbits over time. The behavior of the low-$\alpha$ sequence can be broadly understood in the context of recent work by \cite{ting18}, who explore the observed age dependence of $J_{z}$ for low-$\alpha$ \texttt{APOGEE} RC stars with ages \textless~8 Gyr. \citet{ting18} find that a simple analytic model of vertical heating can describe the trends in the data. They posit that heating is dominated by orbit scattering, presumably from giant molecular clouds (GMCs), with $J_{z}$ $\propto$ $t^{1/2}$. They also take into account how the exponentially declining SFR of the Galaxy results in the decline of GMC occurrence over time, and that the disk mass density decreases with time as well. However, they do not consider high-$\alpha$ sequence stars. We find in Figure \ref{fig:actions} that, unlike the age-$J_{z}$ relationship for the low-$\alpha$ sequence, the high-$\alpha$ age-$J_{z}$ relationship saturates at some stellar age. This saturation matches our physical expectation that, at a certain point, already dynamically hot high-$\alpha$ sequence stars spend such a small fraction of their time near the disk midplane that GMC scattering cannot further heat these stars. 

Given the relationships revealed in Section \ref{sec:age_metallicity} between [Mg/Si] and stellar age, the bottom row of Figure \ref{fig:actions} shows the [Mg/Si] abundance of low-$\alpha$ sequence stars separated into two bins. Divided based on the mean [Mg/Si] value of all the low-$\alpha$ sequence stars, one bin contains the stars with lower [Mg/Si] abundances and the other contains those with higher [Mg/Si] abundances. As seen in the figure, for the radial and vertical actions, at every age there is little to no dynamical separation between stars with low versus high [Mg/Si] enrichment. However for angular momentum, a gradient with [Mg/Si] partly constitutes the scatter in the $J_{\phi}$-age relationship. At every age, stars with lower [Mg/Si] abundances have higher angular momenta than those with higher [Mg/Si] abundances. Due to the large number of stars in the sample, this trend in the mean [Mg/Si] is robust to the standard errors. For the high-$\alpha$ sequence, we find no separation in any action-age relationship with [Mg/Si] abundance. The trends with $J_{\phi}$-[Mg/Si] and age for the low-$\alpha$ sequence reflect the chemical segregation of the star-forming thin disk, at a given time. The absence of a trend with $J_{R}$ and $J_{z}$ suggests that heating and/or migration effects act similarly on mono-age stellar populations at the radii where they reside. For the high-$\alpha$ sequence, the additional absence of a trend with $J_{\phi}$ is reflective of the lack of chemical segregation in the thicker disk in which high-$\alpha$ stars were generally born. These observations support the notion that the low- and high-$\alpha$ sequences experience different modes of enrichment. 

As we have done throughout Section \ref{sec:results}, large samples of stars with high-quality spectra and astrometric measurements allow us to empirically characterize the relationship between chemical abundances, stellar ages, and the dynamical properties of stars. However, the physical origins of these connections between chemistry, age, and dynamics remain an open question in Milky Way science. Nonetheless, empirical investigations can place constraints on theoretical expectations, trends observed in simulations of Milky Way galaxies, and on GCE models. For the remainder of this paper, we now shift our focus and attempt to understand the [Mg/Si] abundance trends we report by connecting the observed abundance patterns to underlying stellar physics.

\section{Chemical evolution modeling}
\label{sec:chempy}

\subsection{Motivation}  
In Section \ref{sec:theory} we established a theoretical motivation for examining the ratio of magnesium to silicon. In Section \ref{sec:extra} we used a large sample of stars with abundance measurements from \texttt{APOGEE} to show that the [Mg/Si] abundance contains different information than the individual [Mg/Fe] and [Si/Fe] abundances. Then, in Section \ref{sec:results} we characterized how the [Mg/Si] abundance varies between the low- and high-$\alpha$ sequences, with age and [Fe/H], with location in the Galaxy, and with the orbital actions. After establishing these empirical [Mg/Si] trends, we now seek to understand these trends by investigating the physical origin of the observed magnesium and silicon abundances.

One way observed abundances can be linked to a physical origin is through GCE modeling. Broadly, the goal of GCE modeling is to employ physically motivated models of star-formation and stellar evolution together with a parameterization of galactic ISM physics to predict observed abundance patterns of stars throughout cosmic time. The simplest form of GCEs involves a single ``zone" that instantiates a site of star-formation surrounded by a reservoir of gas. Through inflow and outflow processes, primordial gas, and gas enriched from the products of star-formation, is exchanged between the gas reservoir and the star-formation site. 

As briefly described in Section \ref{sec:theory}, \textit{Chempy} is a recently developed code that allows for Bayesian inference of GCE model parameters. The publicly available \textit{Chempy} model presented in \cite{chempy} is a one-zone GCE with seven free parameters. This includes three parameters related to stellar physics, and four to the describe the ISM. The stellar physics parameters include the high-mass slope of the IMF, a normalization constant for the number of exploding SN Ia, and the SN Ia enrichment time delay. The ISM parameters include a parameter which sets the star-formation efficiency, the peak of the SFR, the fraction of stellar yields which outflow to the surrounding gas reservoir, and the initial mass of the gas reservoir. With these parameters specifying the GCE, \textit{Chempy} computes the enrichment of the ISM through time based on a set of yield tables. These yield tables describe stellar feedback from three nucleosynthetic channels: CC-SN, SN Ia, and AGB stars. The CC-SN and AGB yields are mass and metallicity dependent, and in the model these yields are ejected immediately following stellar death. SN Ia feedback is independent of mass and metallicity, with the yields being deposited into the ISM according to the \cite{maoz10} DTD. As we will describe in Section \ref{sec:fitting}, with a list of observed stellar abundances, a stellar age estimate, and choice of yield tables, \textit{Chempy} can be used to obtain posterior distributions for the GCE model parameters. 

We present the results of our \textit{Chempy} modeling in the following sections. Our approach is to fit each star's abundance pattern with its own one-zone GCE model, which is in contrast with approaches that fit a single model to either the abundance pattern of multiple stars, or a mean abundance pattern of many stars. However, we caution that ultimately we encounter problems in reproducing the \texttt{APOGEE} abundances, which limits the scope and interpretation of the fits. The challenges we come across are discussed in detail in Section \ref{sec:chempyproblems}, but in short we find that the inferred \textit{Chempy} model parameters result in predicted abundances that are unable to match the observed abundance patterns, particularly the [Si/Fe] abundance. Presumably, this failure is due to the inability of current yield tables to describe the abundance patterns of stars that exhibit a diversity of enrichment histories. 

\subsection{Fitting \texttt{APOGEE} stellar abundances}
\label{sec:fitting}
Given the limitations we encounter when using \textit{Chempy} to model observed \texttt{APOGEE} abundance patterns, we decide to narrow the scope of this study by focusing on just a handful of mono-age, mono-abundance stars. First, we select stars from two narrow [Fe/H] bins: a solar metallicity bin from -0.02 $\textless$ [Fe/H] $\textless$ 0.02 dex, and a metal-poor bin from -0.32 $\textless$ [Fe/H] $\textless$ -0.28 dex. To consider trends with age, we further select stars in three age bins:  1.5 - 2.5 Gyr, 4.5 - 5.5 Gyr, and 8.5 - 9.5 Gyr. We then randomly select 10 stars in each [Fe/H] and age bin, with five stars having a high probability of belonging to the low-$\alpha$ sequence, $P(z$ = high-$\alpha$) $\textless$ 0.5, and the other five stars having a high probability of belonging to the high-$\alpha$ sequence, $P(z$ = high-$\alpha$) $\textgreater$ 0.5.

We use the default \textit{Chempy} yield tables which include CC-SN yields from \cite{nomoto13}, SN Ia yields from \cite{seitenzahl13}, and AGB yields from \cite{karakas10}. Based on these yield tables, the \textit{Chempy} inference routine can fit a list of 22 chemical abundances. However, since the focus in this paper is on quantifying the information encoded in magnesium and silicon abundances, we fit just these two abundances, as well as [Fe/H]. We also pass the age estimates from \cite{ness16} to \textit{Chempy} to be used during inference.

While the \textit{Chempy} GCE model is described by seven parameters, we decide to only infer four of them. This is so the posterior is not overwhelmingly dominated by the priors, as the likelihood is only being computed from three abundances. The three we decide not to infer are the SN Ia delay time, the fraction of stellar feedback that returns to the surrounding gas reservoir, and the initial mass of gas reservoir. Through empirical experimentation, we notice our data to be relatively uninformative regarding these parameters. Additionally, as described in \cite{chempy}, observational constraints on these three parameters are less certain. During the Markov chain Monte Carlo (MCMC) routine, the parameters that we do not infer are set to default priors values as given in \cite{chempy}.

The four \textit{Chempy} parameters we infer are described in detail as follows. Two of them, $\alpha_{\rm IMF}$ and $\mathrm{N}_\mathrm{Ia}$, are typical SSP parameters. The IMF parameter is defined as the high-mass slope of a Chabrier IMF \citep{chabrier03}, which in the \textit{Chempy} model is defined over the stellar mass range of 0.08 - 100 M$_{\odot}$. The IMF parameter determines the ratio of low- versus high-mass stars formed in a burst of star-formation, which alters the relative yields produced. The SN Ia parameter, $\mathrm{N}_\mathrm{Ia}$, is the normalization constant of the \cite{maoz10} DTD, in which SN Ia enrichment is modeled as a power law in time. In \textit{Chempy} specifically, $\mathrm{N}_\mathrm{Ia}$ is defined as the number of SNIa explosions per solar mass over a 15 Gyr time period. The remaining two parameters we infer are ISM parameters, star formation efficiency (SFE) and $\mathrm{SFR}_\mathrm{peak}$. The SFE governs the amount of gas infalling from the gas reservoir that is needed to sustain the conversion of ISM gas to stars. This is defined as $m_{\rm SFR}$/$m_{\rm ISM}$. As illustrated in Figure 7 of \cite{chempy}, low SFE values result in more gas being required for the star formation. That leads to a bigger gas reservoir in the ISM, which dilutes the stellar enrichment and keeps the ISM metallicity lower compared to a higher SFE value. The fourth parameter we infer is the peak time of the SFR, $\mathrm{SFR}_\mathrm{peak}$. In \textit{Chempy} this is parameterized as a gamma distribution with $\mathrm{SFR}_\mathrm{peak}$ as the scale parameter, and the shape parameter fixed at $k=2$. As seen in Figure 6 of \cite{chempy}, an early SFR peak results in star-formation that is concentrated at early times, whereas a later SFR peak results in a smoother SFR distribution. 

We adopt the same priors for all of the stars that we fit. Each prior is defined as a normal distribution with mean $\mu$ and width $\sigma$, as listed in Table \ref{tab:params}. Here we report both the priors used in \cite{chempy}, and the priors we choose to adopt in this work. The priors set forth in \cite{chempy} are motivated by recent work related to observational estimates of Milky Way-like GCE parameters. While adopting these priors in the \textit{Chempy} model results in agreement with the Sun's chemistry, for our work we choose to broaden the priors. This is because we are fitting stars of various ages that exhibit a range of chemical abundance patterns, which is tentatively reflective of a multitude of enrichment histories. We increase the width of the priors to better capture the uncertainty in our knowledge regarding the parameter values that may describe this diversity of stars. 

\begin{table}
\begin{centering}
\tabletypesize{\scriptsize}
\caption{\textit{Chempy} parameter priors}
\label{tab:params}
\tablewidth{0pt}

\begin{tabular}{rccccccc}

\hline
Parameter Name & \multicolumn{3}{c}{\textit{Chempy} Default} &  & \multicolumn{3}{c}{This Work}  \\ 

\hline
\hline
& $\mu$ & & $\sigma$ & \vline & $\mu$ & & $\sigma$ \\ \cline{2-4} \cline{5-8}

$\alpha_\mathrm{IMF}$ \vline & -2.29 & $\pm$ & 0.2 & \vline & -2.29 & $\pm$ & 0.3 \\
$\log_{10}\left(\mathrm{N}_\mathrm{Ia}\right)$ \vline & -2.75 & $\pm$ & 0.3 & \vline & -2.75 & $\pm$ & 0.5  \\
$\log_{10}\left(\mathrm{SFE}\right)$ \vline & -0.3 & $\pm$ & 0.3 & \vline & -0.3 & $\pm$ & 0.5 \\
$\log_{10}\left(\mathrm{SFR}_\mathrm{peak}\right)$\footnote{In \cite{chempy}, a Gaussian prior in linear space with $\mu$ = 3.5 Gyr and $\sigma$ = 1.5 Gyr is placed on the SFR$_{\rm peak}$ parameter.} \vline & 0.55 & $\pm$ & 0.1 & \vline & 0.45 & $\pm$ & 0.3 \\

\hline
\hline
\vspace{-2ex}


\end{tabular}

\end{centering}
\end{table}

As described in \cite{chempy}, inference of the \textit{Chempy} GCE parameters is achieved through posterior sampling, specifically utilizing the \texttt{emcee} implementation of MCMC \citep{emcee}; \texttt{emcee} is an affine-invariant ensemble sampler, which initializes a collection of walkers in a defined parameter space. Each walker evaluates the posterior at every step of its random walk through the space, and by comparing the current evaluation to the evaluation at the previous step, the proposed step is either accepted or rejected with some probability. In the context of \textit{Chempy}, the sampling routine is summarized as follows. During each walker's steps, the current parameter configuration is used to evaluate the log prior. Then, keeping track of the chemical abundances at each timestep, the \textit{Chempy} GCE model is run with these parameter values from $t$ = 0 to the provided estimated time of stellar birth. The log likelihood is then computed by summing the square difference between the observed and model abundances, weighted by the observational uncertainties. The posterior is then evaluated by summing the log prior and log likelihood.  

\begin{figure}[tp]
\centering
\includegraphics[width=1\columnwidth]{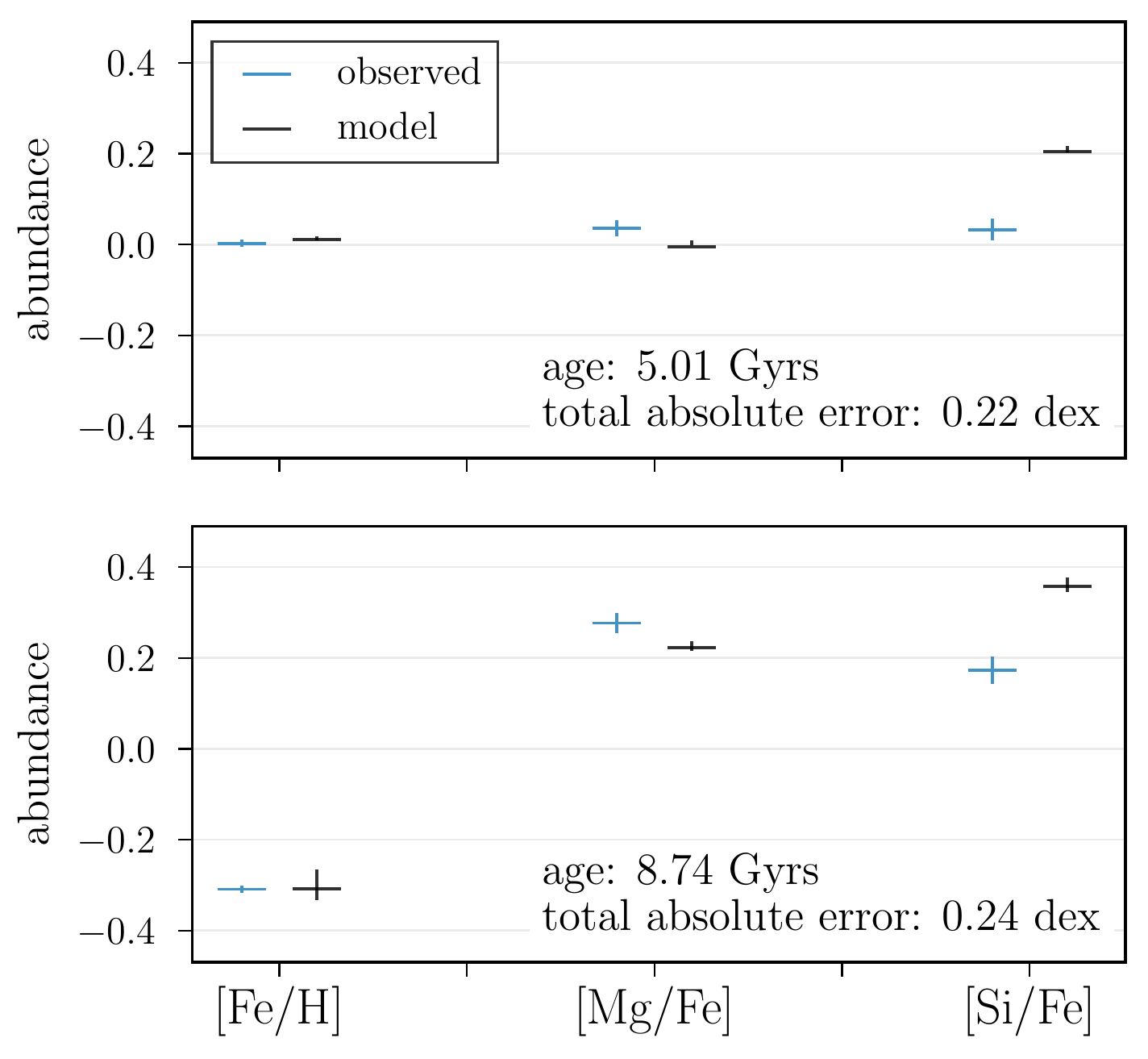}
\caption{Comparison of the observed abundances to the \textit{Chempy} predicted abundances for two example stars. The uncertainties on the observed abundances are indicated, and for the predicted abundances the range of possible abundance values obtained from the 1$\sigma$ posterior distribution is shown.}
\label{fig:ppv}
\end{figure}

For each of the stars we fit, we initialize the MCMC routine with 28 walkers. The starting locations of the walkers are typically set to be confined to a tight 4-dimensional ball centered on the prior means, and after several iterations the walkers explore the larger, relevant parameter space. Instead of running each MCMC instance for a pre-defined number of steps, we monitor the convergence of the chains in real time and terminate the runs once some convergence criterion is satisfied. The commonly used Gelman-Rubin statistic is not appropriate for chains produced with \texttt{emcee} since the chains are not independent of one another. Instead, \cite{emcee} suggest assessing convergence by computing the integrated autocorrelation time of the chains, $\tau$, which is an estimate of the number of steps the walkers need to draw an independent sample from the posterior distribution. Chains with longer estimated $\tau$'s require more steps to generate a given number of independent posterior samples. We modify the \textit{Chempy} inference routine to compute an estimate of $\tau$ for each chain every 300 steps. We do this using the \texttt{emcee} implementation of \cite{goodman10}'s method for estimating the integrated autocorrelation time. Once the chains reach a length of 30$\times\tau$, they are terminated. Figure \ref{fig:convergence} in the Appendix shows the distribution of integrated autocorrelation times for each of the runs, as well as the estimated number of independent posterior samples generated. Often, more than 30 independent samples are obtained because longer chains are needed to generate a reliable estimate of $\tau$. In addition to being used to monitor convergence, we utilize the integrated autocorrelation times to remove the burn-in phase of the chains. For each star, we determine the maximum integrated autocorrelation time across all parameters and chains, and remove three times this value before computing the posterior percentiles.

\subsection{Results}
\label{sec:fit_results}
We now present the results of the \textit{Chempy} GCE modeling for the mono-age, mono-abundance stellar samples described above. As an example of the full posterior distributions obtained, Figure \ref{fig:corner} in the Appendix shows the trace plots and joint posterior distributions for two stars that we fit: a solar-metallicity, 5 Gyr-aged star in the low-$\alpha$ sequence, and a metal-poor, 9 Gyr-aged star in the high-$\alpha$ sequence. Details regarding these particular fits are discussed in the Appendix. While there are potentially interesting differences between the posteriors of the two stars, these differences must be interpreted in the context of how well the models describe the data. To examine this we use the inferred parameter values to run instances of the \textit{Chempy} GCE model and predict the stellar abundance patterns at the time of stellar birth. We then compare these predicted abundances to the observed abundances that were used to infer the posteriors. This approach to model evaluation is similar to posterior predictive checking \citep{gelman}. Figure \ref{fig:ppv} shows the observed and predicted abundances for both of the stars included in Figure \ref{fig:corner}. The observed abundances are the [Fe/H], [Mg/Fe], and [Si/Fe] measurements from \texttt{APOGEE} that were passed to the likelihood calculation, and the predicted abundances are based on a \textit{Chempy} model run using parameter values from each star's posterior distributions. The horizontal markers indicate predicted abundances based on the 50$^{\rm th}$ percentile posterior values, while the vertical range indicates the predicted abundance range considering \textit{Chempy} models with the 16$^{\rm th}$ and 84$^{\rm th}$ percentile values of the parameter posteriors.

\begin{figure*}[tp]
\centering
\includegraphics[width=2\columnwidth]{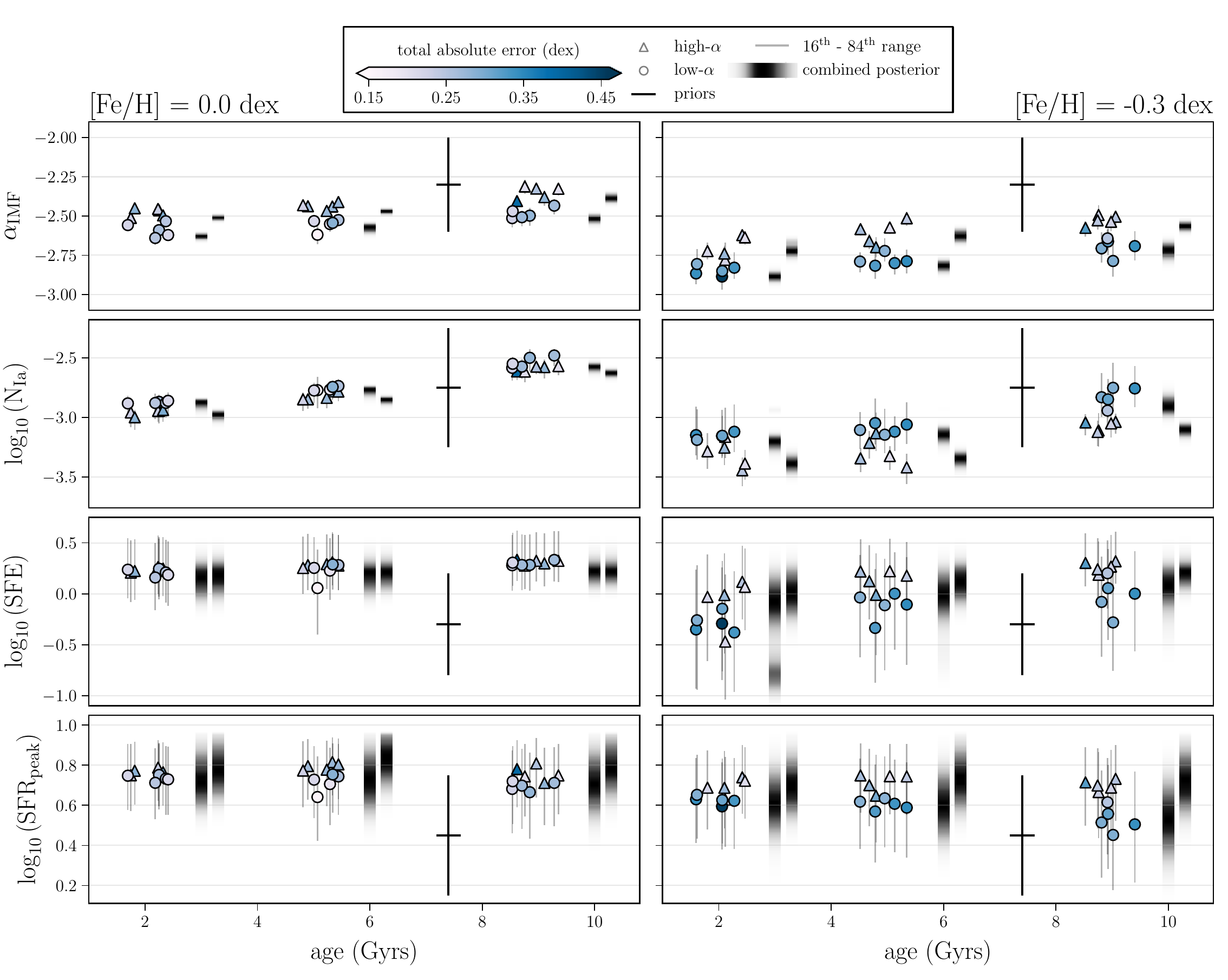}
\caption{\textit{Chempy} posterior distributions for mono-age, mono-abundance stars with abundance measurements from \texttt{APOGEE}. Each panel shows the values of the inferred GCE parameters ($\alpha_{\rm IMF}$, N$_{\rm Ia}$, SFE, and SFR$_{\rm peak}$) for stars with ages of 2, 5, and 9 Gyr. The plots on the left show stars with solar [Fe/H] abundances, while the plots on the right show stars with [Fe/H] abundances of -0.3 dex. For each star, the 50$^{\rm th}$ percentile of the posterior distribution is indicated by the markers, where high-$\alpha$ sequence stars are indicated by triangles and low-$\alpha$ sequence are indicated by circles. The error bars on the markers span the 1$\sigma$ (or 16$^{\rm th}$ - 84$^{\rm th}$) posterior percentile range, and the markers are colored by the total absolute error between the observed and predicted abundances. The prior on each parameter is shown for comparison, with the variance of the distribution indicated by the solid black line. The combined posteriors, binning stars with the same age, [Fe/H] abundance, and $\alpha$-enhancement, are shown in grayscale, with darker shades indicating a higher posterior probability. }
\label{fig:fits}
\end{figure*}

As seen in the figure, the [Fe/H] and [Mg/Fe] abundances of the 5 Gyr-aged solar-metallicity star are well predicted by the model, falling nearly within the observational uncertainty ranges. However, the [Si/Fe] abundance is not reproduced as well. The model over-predicts the silicon abundance by nearly 0.2 dex, and the observed abundance is not within the model's 1$\sigma$ posterior range. The predicted abundances for the 9 Gyr-old metal-poor star exhibit similar discrepancies with respect to the observed abundances. As seen in the \textit{bottom panel} of Figure \ref{fig:ppv}, the \textit{Chempy} model run with the 50$^{\rm th}$ percentile posterior values results in predicted abundances that agree or nearly agree with the observed [Fe/H] and [Mg/Si] abundances. But, similar to the solar-metallicity star, the range of predicted [Si/Fe] abundance is disparate from the observed [Si/Fe] abundance, and is over-predicted by about 0.2 dex. 

The main takeaway from Figure \ref{fig:ppv} is as follows. While we are relatively confident that the chains are not unconverged, we find that the inferred parameters do not always result in \textit{Chempy} models that are able to reproduce the observed abundances, particularly when it comes to [Si/Fe]. Possible reasons for this will be discussed further in Section \ref{sec:chempyproblems}. While we next present how the inferred GCE parameters vary for stars of different ages and abundances, the quality of the model fits cautions any significant physical interpretation of these trends. 

Moving beyond the two example stars discussed up to this point, we now present the posteriors for the full sample of mono-age, mono-abundance stars that we fit. As discussed in Section \ref{sec:fitting}, we fit a collection of 60 stars that fall into two [Fe/H] bins, three age bins, and have $\alpha$-enhancements characteristic of both the low- and high-$\alpha$ sequences. Figure \ref{fig:fits}, which displays summary statistics of the posterior distributions at each age and [Fe/H], summarizes the results of these fits. In each panel, the individual posteriors for each star are described by the markers placed at the 50$^{\rm th}$ percentile value, and the error bars which span the 1$\sigma$ (or 16$^{\rm th}$ - 84$^{\rm th}$) percentile range. The markers are colored by the ``total absolute error", which is the sum of the absolute difference between the observed abundances and predicted abundances based on the 50$^{\rm th}$ posterior values. We use this error as a goodness-of-fit measure, which we discuss further in Section \ref{sec:chempyproblems}. We also show the combined posterior, which is indicated by the grayscale colorbar. These combined posteriors are determined by multiplying the individual posterior distributions of all the stars that fall within the same [Fe/H], age, and $\alpha$ sequence bin, which assumes that individual posterior probability distributions are independent of one another. Given that the stars have similar properties modulo any differences in [Mg/Fe] and [Si/Fe], the combined posteriors more strongly constrain the parameters than the individual posteriors. 

First consider the $\alpha_{\rm IMF}$ parameter, which is the GCE parameter we find to be most constrained by the data. For all the stars that we fit, the posteriors appear to disfavor IMF slopes that are more top-heavy than $\alpha_{\rm IMF}$ \textgreater~-2.25. Additionally, for both the solar-metallicity and metal-poor stars, we find a weak trend between the inferred $\alpha_{\rm IMF}$ and stellar age. Younger stars are found to have IMF slopes described by a more bottom-heavy IMF, whereas older stars are described by more Chabrier-like IMF slopes. At each age, there is a further trend with $\alpha_{\rm IMF}$ and [Fe/H]. Metal-poor stars are found to be described by a more bottom-heavy IMF slope than their solar-metallicity counterparts of the same age. There are also smaller differences between the inferred IMF slope of low- and high-$\alpha$ sequence stars, which is most apparent when considering the combined posteriors. For both metallicities and at all ages, high-$\alpha$ sequence stars are found to have more top-heavy IMF slopes than low-$\alpha$ sequence stars. 

Aside from the IMF parameter, N$_{\rm Ia}$ is the other parameter we find to be tightly constrained by the \texttt{APOGEE} abundances. As seen in the second row of Figure \ref{fig:fits}, the $\sim$1$\sigma$ posterior ranges appear to disfavor different regions of the N$_{\rm Ia}$ parameter space, with dependence on stellar age and metallicity. First we notice that at every age, the metal-poor stars are found to be fit by fewer SN Ia than their solar-metallicity counterparts of the same age. This difference is $\sim$0.25 dex on average. In addition to the variations with metallicity, there are also trends with age at fixed [Fe/H]. For the solar-metallicity stars, younger stars are found to be described by fewer SN Ia compared to older stars, spanning values from log(N$_{\rm Ia}$) = -3 to -2.5. For the metal-poor stars, a similar trend is observed with age, except these values span log(N$_{\rm Ia}$) = -3.5 to -2.75. Lastly, similar to the IMF parameter, there are small differences between the N$_{\rm Ia}$ posteriors of high-$\alpha$ sequence stars and low-$\alpha$ sequence stars. As apparent in the combined posteriors, low-$\alpha$ sequence stars are described by fewer SN Ia than high-$\alpha$ sequence stars, at every age and metallicity. 

Unlike the IMF and SN Ia parameters, the two ISM parameters are not as tightly constrained by the data. As seen in the third row of Figure \ref{fig:fits}, the 16$^{\rm th}$ - 84$^{\rm th}$ ranges of the SFE posteriors are more comparable to the range of the prior, especially for the metal-poor stars. The binned posteriors of metal-poor stars are more constraining, and show the SFE values for these stars to be $\textgreater$ -0.25 dex, with weak trends seen with age and $\alpha$-enhancement. For the 2 Gyr-aged metal-poor stars, we see a discrepancy between the 50$^{\rm th}$ percentile values of the individual posteriors and the combined posterior distributions, and bimodality in the combined posterior for the high-$\alpha$ sequence. This is a result of these posteriors being non-Gaussian, exhibiting a one-sided tail in addition to a peak. For the solar-metallicity stars, the posteriors are more constrained. For these stars, SFE values less than $\sim$0 dex are mostly disfavored, with posterior medians at $\sim$0.25 dex for stars of each age and $\alpha$ enhancement. Compared to stars of the same age that are metal-poor, the solar-metallicity stars exhibit slightly higher SFE values.

Finally, the fourth row of Figure \ref{fig:fits} shows the posteriors for SFR$_{\mathrm{peak}}$. As with the SFE parameter, the ranges of the SFR$_{\rm peak}$ posteriors are more comparable to the range of the prior. Despite this, some regions of the parameter space are still disfavored. For the solar-metallicity stars, their posteriors suggest a low probability of an SFR peak occurring before $\sim$3 Gyr, with a median posterior value of $\sim$5.5 Gyr for all of the stars, with only a slight difference between the low-$\alpha$ and high-$\alpha$ sequence. For the metal-poor stars, their posteriors suggest a low probability of an SFR peak occurring before $\sim$2 Gyr, with the median posterior value of $\sim$4 Gyr, with a slight trend with age and $\alpha$ enhancement.

\subsection{Limitations}
\label{sec:chempyproblems}
As emphasized throughout Section \ref{sec:fit_results}, we find that the predicted \textit{Chempy} abundances are often in tension with the observed abundances, particularly with regard to [Si/Fe]. We now quantify these discrepancies, and discuss possible reasons for them that ultimately limit the interpretation of our chemical evolution modeling results. In the end, we advocate that, given the current quality and quantity of abundance measurements for stars located all throughout the Milky Way, a more data-driven approach to GCE modeling is a possible path forward. 

Before examining how well the \textit{Chempy} models are able to reproduce the abundance patterns of stars with varying metallicities, ages, and $\alpha$-enhancements, we first compare the default and alternative \textit{Chempy} yield tables. As discussed in \cite{chempy}, the choice of yield tables is a hyperparameter of the \textit{Chempy} model, and ideally the observed abundances can be used to determine which yield tables are most probable. Consider the posteriors inferred under the default yield tables (based on \cite{nomoto13}, \cite{seitenzahl13}, and \cite{karakas10}), which are the tables used for the results shown in Section \ref{sec:fit_results}. 

For the solar-metallicity stars, the average absolute difference between the predicted and observed [Fe/H] abundances is 0.007 dex, with all of the stars having their abundances over-predicted. For [Mg/Fe] the predicted abundances match the observed abundances somewhat worse, with the average absolute difference being 0.056 dex (100\% under-predicted). However, the \textit{Chempy} models are unable to reproduce the observed silicon abundances. The average absolute difference between the predicted and observed [Si/Fe] abundances is 0.187 dex (100\% over-predicted). For the metal-poor stars, the predicted abundances are less consistent with the observed abundances. On average, the [Fe/H] predictions are 0.035 dex off from the observed abundances (100\% over-predicted), the [Mg/Fe] predictions are 0.073 dex off from the observed abundances (100\% under-predicted), and the [Si/Fe] predictions are 0.178 dex off from the observed abundances (100\% over-predicted). 

The differences between the [Fe/H] and [Mg/Fe] predictions are at least partly due to the smaller measurement uncertainties on the [Fe/H] abundances. These smaller [Fe/H] uncertainties result in larger penalties in the likelihood evaluation for discrepant [Fe/H] predictions compared to the penalties for discrepant [Mg/Fe] predictions. The observational uncertainties on the [Mg/Fe] and [Si/Fe] abundances are comparable, so the observational uncertainties are not the primary driver of the [Si/Fe] over-prediction.

To test if perhaps another set of yield tables could result in better [Si/Fe] predictions, we also infer the parameter posteriors assuming the alternative \textit{Chempy} yield tables (based on \cite{chieffi04}, \cite{thielemann03}, and \cite{ventura13}). For the solar-metallicity stars, the predicted [Fe/H] abundances are off by 0.06 dex (96\% under-predicted, 4\% over-predicted), the predicted [Mg/Fe] abundances are off by 0.11 dex (100\% over-predicted), and the predicted [Si/Fe] abundances are off by 0.32 dex (100\% over-predicted). Compared to the default yield tables, the alternative yields result in poorer abundance matches. Considering the metal-poor stars, the predicted [Fe/H] abundances are off by 0.09 dex (100\% under-predicted), the predicted [Mg/Fe] abundances are off by 0.14 dex (100\% over-predicted), and the predicted [Si/Fe] abundances are off by 0.34 dex (100\% over-predicted). Again, the alternative yield tables result in worse matches to the observed abundances.

We make several observations from this comparison of the predicted and observed abundances for both the default and alternative yield tables. First is that the default yield tables result in better matches to the observed abundances for all elements, for both the solar-metallicity and metal-poor stars. As found in \cite{P18} who present a scoring system for comparing yield tables, the optimal choice of tables even when fitting only proto-solar abundances is dependent on what specific abundances are being fit. Presumably the optimal yield tables will also differ for stars with varying abundance patterns. A detailed ranking of the various yield tables for a collection of stars with different abundance properties may reveal informative trends regarding what tables are preferred by the data. The other observation we make is that, regardless of yield table choice, the silicon abundance is the most difficult to reproduce and is systematically over-predicted by $\sim$0.2 - 0.4 dex. Because of this, despite the [Mg/Si] trends that the default \textit{Chempy} yield tables exhibit in Section \ref{sec:theory}, the inferred parameter values are not able to correctly reproduce the relative amounts of magnesium and silicon. As found in \cite{P18} there seems to be an Mg to Si offset for all tested CC-SN yield tables, which include those used in this work and more recent ones from the literature; see their Figure 4. It is unlikely that other nucleosynthetic channels or the GCE model parameterization are causing this offset. Instead, it is found that the explosion energies of the CC-SN could be the reason and lowering them could remedy the discrepancy \citep{H10, F18}. 

While the default yield tables are preferred by the data, these yield tables result in some systematic trends between stellar properties and how well the \textit{Chempy} models can reproduce the abundance patterns of stars (as seen in Figure \ref{fig:fits}). While we find no significant trends with stellar age, there are some differences in how well the models describe metal-poor versus solar-metallicity stars, and low-$\alpha$ versus high-$\alpha$ sequence stars. Considering metallicity, we find that the abundances of the solar metallicity stars are reproduced slightly better than those of the metal-poor stars. To compare the fits, we compute the ``total absolute error" between the predicted and observed abundances by summing the absolute values of the individual [Fe/H], [Mg/Fe], and [Si/Fe] differences. The average total error of the solar-metallicity stars is 0.25 dex, while that of the metal-poor stars is 0.29 dex. As mentioned previously, a majority of this error is from the model's over-prediction of silicon. The worse model predictions for metal-poor stars compared to solar-metallicity stars could be due to a number of factors, and we note that a similar finding is reported in \cite{chempy}, where Arcturus with an [Fe/H] $\sim$ -0.5 dex is fit more poorly than the Sun. Generally, stellar models are gauged to solar abundances, which could result in the yields preferentially producing solar abundance patterns.

We also find differences in the ability of the models to reproduce the abundances of low- versus high-$\alpha$ sequence stars. First considering solar-metallicity stars of all ages, we find that on average the high-$\alpha$ stars have marginally higher total errors than low-$\alpha$ stars (0.26 dex compared to 0.24 dex). This small difference between the two sequences is mostly due to worse [Mg/Fe] predictions for the high-$\alpha$ stars. For the metal-poor stars, we find a larger discrepancy in the accuracy of the predicted abundances between the low- versus high-$\alpha$ sequences. On average the high-$\alpha$ stars have a total error of 0.25 dex, while the low-$\alpha$ stars have an average total error of 0.32 dex. This difference is mainly from both the [Fe/H] and [Si/Fe] abundance predictions, which are each $\sim$0.03 dex worse for the low-$\alpha$ sequence stars. The more significant difference between the low- and high-$\alpha$ sequence fits at low metallicities could be a consequence of the high-$\alpha$ sequence abundance pattern resulting directly from CC-SN yields. Fine-tuning of the abundances with Fe from SN Ia is additionally needed to produce low-$\alpha$ sequence abundance patterns.

In summary, while we find the yield tables of \cite{nomoto13}, \cite{seitenzahl13}, and \cite{karakas10} to be preferred by the data, ultimately we are not able to describe the abundance patterns of these \texttt{APOGEE} stars due to the systematic over-prediction of [Si/Fe]. The reasons for this fall into three categories, which include shortcomings with the yield tables, the model, or the data. Considering the yield tables, one possibility is that there are nucleosynthetic contributions missing from the yield tables that are necessary to reproduce the Si abundances of these stars. The yield tables might also not be descriptive enough to accurately characterize stars with abundance patterns arising from a variety of initial environmental conditions and that have experienced a multitude of different evolutionary pathways. On the model side, while here we fit each star with its own unique one-zone model, the ISM parameterization might be too simplistic to reproduce the abundance patterns of these stars. Parameterizations that allow for things like a bursty star-formation history, an adaptive coronal gas mass and volume, or different mixing of the ISM will better describe the abundances. Finally, shortcomings may lie with the data. In this paper, our primary focus was to empirically characterize the ratio of magnesium to silicon, and then understand the physical origin of these variations. With this goal, we fit \textit{Chempy} with just three abundances ([Fe/H], [Mg/Fe], and [Si/Fe]) in an attempt to generate the simplest versions of the model that would be able to reproduce the data. However, GCE models could potentially be better constrained by different or more abundances, or even the entire chemical abundance vector available from \texttt{APOGEE}. That said, two abundances are highly constraining in the sense that they highlight deepened tensions between the data, yield tables, and GCE models, even in a simple application.

\section{Summary and Conclusion}
\label{sec:conclusion}
In this paper we present a detailed investigation of the [Mg/Si] abundance in the Milky Way disk. We do this using a large sample of stars with \texttt{APOGEE} abundance measurements, estimated stellar ages, and \textit{Gaia} astrometry. Inspired by the increasing precision and sheer number of stars with available abundances, our primary goal is to go beyond the information contained in a bulk $\alpha$ abundance and examine the higher level of granularity encoded in an inter-family abundance ratio. The specific choice of magnesium and silicon is motivated by the expected subtle differences in their nucleosynthetic origins, which has been previously used to interpret the [Mg/Si] abundances of stars in the Sattitargius dwarf galaxy. Our endeavor for Milky Way stars includes both an empirical characterization of the [Mg/Si] abundance throughout the Galaxy, as well an attempt to link [Mg/Si] variations to a physical origin through galactic chemical evolution modeling. 

After gaining intuition in Section \ref{sec:theory} for how Mg and Si yields evolve through time in a single burst of star-formation, we then focus on establishing how the observed [Mg/Si] abundance varies in the Galaxy. We make connections between [Mg/Si] and various stellar properties, including stellar age, [Fe/H], location, and stellar orbital actions. With the goal of better understanding the origin of the Milky Way's bimodal $\alpha$ sequence, we identify differences in [Mg/Si] between low- and high-$\alpha$ sequence stars. Our findings are summarized as follows.

\begin{itemize}
\item High-$\alpha$ sequence stars are more enhanced in [Mg/Si] than low-$\alpha$ sequence stars, with the difference in the average [Mg/Si] between the two sequences being $\sim$0.08 dex. The two sequences also exhibit distinct behavior in the variation of their [Mg/Si] abundances across the [$\alpha$/Fe]-[Fe/H] plane, where [Mg/Si] varies more strongly with [$\alpha$/Fe] and [Fe/H] in the low-$\alpha$ sequence than it does in the high-$\alpha$ sequence. Given the expected theoretical Mg and Si yields (Figure \ref{fig:theory_mgsi}), observed variations in [Mg/Si] at early times could potentially be used to discriminate between different IMFs. This is tentatively supported by the \textit{Chempy} inferences in Section \ref{sec:chempy}, where high-$\alpha$ sequence stars are found to be described by a slightly more top-heavy IMF than low-$\alpha$ sequence stars. 

\item The enhanced [Mg/Si] abundance of the high-$\alpha$ sequence compared to the low-$\alpha$ sequence is persistent at all stellar ages. Considering the evolution of [Mg/Si] with stellar age, from old to young ages the [Mg/Si] abundance of the low-$\alpha$ sequence decreases nearly 2$\times$ more than the [Mg/Si] abundance of the high-$\alpha$ sequence. This difference is persistent for solar-metallicity stars, as well as metal-poor stars, but is minimized at higher [Fe/H] abundances where the two $\alpha$ sequences become less distinct. 

\item Inspired by \cite{hayden15}'s characterization of the [$\alpha$/Fe]-[Fe/H] distribution in the disk, we examine how the [Mg/Si]-[Fe/H] distribution varies with location in the Galaxy, from R = 3 - 15 kpc and $|z|$ = 0 - 2 kpc. We find that the [Mg/Si]-[Fe/H] distributions of the high- and low-$\alpha$ sequences considerably overlap at each R and $|z|$, but that there are significant differences in how the mean [Mg/Si] of the two sequences varies throughout the Galaxy. For the high-$\alpha$ sequence, the trend with mean [Mg/Si] and radius is the same at all heights from the disk midplane, with the highest [Mg/Si] occurring at R $\sim$ 8 kpc. However, for the low-$\alpha$ sequence the highest [Mg/Si] abundances occur at R $\sim$ 6 kpc and the mean [Mg/Si] at each radius scales with $|z|$. Since the [Mg/Si] abundance is correlated with age, these trends seemingly reflect the age gradients with R and $|z|$ across the disk, which are distinct for the low- and high-$\alpha$ sequences. 

\item Considering trends with orbital actions, our findings confirm the results of \cite{gandhi} that the high- and low-$\alpha$ sequences are distinct in $J_{\phi}, J_{r}$, and $J_{z}$ at all stellar ages. Examining how [Mg/Si] varies with actions, we find that at all ages low-$\alpha$ sequence stars with lower [Mg/Si] abundances have higher angular momenta than their enriched [Mg/Si] counterparts. A similar trend is not found for the radial or vertical actions, or the high-$\alpha$ sequence and any action. Presumably, this reflects the radial metallicity gradient in the disk at a given age, and a bulk chemical composition of star-forming gas that varies with time and Galactocentric radius, but not with distance from the disk midplane. 
\end{itemize}

\noindent These observed trends with [Mg/Si] abundance support established and reveal new connections between chemistry and orbital dynamics. Given the expected differences in the chemical enrichment processes responsible for generating Mg and Si yields, the varying [Mg/Si] abundance relations between low versus high-$\alpha$ sequence stars bolsters the notion that the two sequences are distinct in their origin and formation. Detailed matching of these observational trends between [Mg/Si], age, and dynamics to the properties of simulated Milky Way-like galaxies can place powerful constraints on disk formation and chemical evolution mechanisms. 

In the second half of this paper we focus our attention on understanding the physical origin of the observed [Mg/Si] variations. Specifically, we perform GCE modeling, which combines physically motivated models of star-formation, the ISM, and galactic evolution to predict the chemical content of the ISM through time. Here, we employ the recently developed \textit{Chempy} code which, given a set of observed abundances and estimated stellar age, allows for inference of GCE model parameters. 

We infer GCE parameters for a set of mono-age, mono-metallicity stars with various [Mg/Fe] and [Si/Fe] abundances. For each star, we obtain posterior distributions for four parameters: the slope of the high-mass IMF, the number of SN Ia explosions per solar mass over a 15 Gyr time period, the SFE, and the peak of the SFR. While there are potentially interesting relationships between these parameters and stellar properties, our interpretation of them is limited. This is because the \textit{Chempy} models are unable to reproduce the \texttt{APOGEE} abundances for the diversity of stars that we fit. A majority of the discrepancy between the predicted and observed abundances comes from the consistent over-prediction of [Si/Fe], which has also been reported by \cite{P18} for a number of tested CC-SN yield tables. We also find systematic trends with metallicity and $\alpha$ enhancement, and the predictive quality of the \textit{Chempy} models. This reveals tensions between the GCE models, yield tables, and observed abundance measurements. 

The main conclusions of this paper are as follows. Given the large number of stars with high-quality abundance measurements available, small variations in these abundances can be characterized by averaging stars with similar properties, such as $\alpha$-enhancement, location, age, or actions. In this way, we were specifically motivated to examine the inter-family ratio of two $\alpha$-elements, Mg and Si, because of expected differences in how they are produced in CC-SN and SN Ia events. This approach we take in dissecting the [Mg/Si] abundance can be generalized to isolate other particular enrichment channels. In theory, other inter-family or intra-family abundances can be used to probe specific nucleosynthesis or disk formation mechanisms, and these empirical relationships can serve as detailed constraints for simulations of Milky Way-like galaxies. However, in practice we encountered challenges in connecting variations in Mg and Si to underlying stellar physics. Specifically, we found that a flexible model of GCE is unable to predict even just three stellar abundances that represent a diversity of enrichment histories. This failure highlights tensions between the chemical evolution models, the yield tables, and the data. As we have now entered a regime of rich stellar abundance information, a more data-driven approach to chemical evolution models and nucleosynthetic yield tables may be a way forward.

\section*{Acknowledgements}
We would like to thank Gail Zasowski, Dan Foreman-Mackey, Victor Debattista, and Adam Wheeler for helpful conversations.  We are also grateful to Nick Carriero for his assistance in planning the \textit{Chempy} runs on the Flatiron Institute computing cluster.

KB is supported by the NSF Graduate Research Fellowship under grant number DGE 16-44869. KB thanks the LSSTC Data Science Fellowship Program, her time as a Fellow has benefited this work. MN is supported by the Alfred P. Sloan Foundation. KVJ's contributions were supported in part by the National Science Foundation under grants NSF PHY-1748958 and NSF AST-1715582. Her work was performed in part during the Gaia19 workshop and the 2019 Santa Barbara \textit{Gaia} Sprint (also supported by the Heising-Simons Foundation), both hosted by the Kavli Institute for Theoretical Physics at the University of California, Santa Barbara.

Funding for the Sloan Digital Sky Survey IV has been provided by the Alfred P. Sloan Foundation, the U.S. Department of Energy Office of Science, and the Participating Institutions. SDSS-IV acknowledges
support and resources from the Center for High-Performance Computing at
the University of Utah. The SDSS website is www.sdss.org. SDSS-IV is managed by the Astrophysical Research Consortium for the 
Participating Institutions of the SDSS Collaboration including the 
Brazilian Participation Group, the Carnegie Institution for Science, 
Carnegie Mellon University, the Chilean Participation Group, the French Participation Group, Harvard-Smithsonian Center for Astrophysics, 
Instituto de Astrof\'isica de Canarias, The Johns Hopkins University, 
Kavli Institute for the Physics and Mathematics of the Universe (IPMU) / 
University of Tokyo, Lawrence Berkeley National Laboratory, 
Leibniz Institut f\"ur Astrophysik Potsdam (AIP),  
Max-Planck-Institut f\"ur Astronomie (MPIA Heidelberg), 
Max-Planck-Institut f\"ur Astrophysik (MPA Garching), 
Max-Planck-Institut f\"ur Extraterrestrische Physik (MPE), 
National Astronomical Observatories of China, New Mexico State University, 
New York University, University of Notre Dame, 
Observat\'ario Nacional / MCTI, The Ohio State University, 
Pennsylvania State University, Shanghai Astronomical Observatory, 
United Kingdom Participation Group,
Universidad Nacional Aut\'onoma de M\'exico, University of Arizona, 
University of Colorado Boulder, University of Oxford, University of Portsmouth, 
University of Utah, University of Virginia, University of Washington, University of Wisconsin, 
Vanderbilt University, and Yale University. \\

\textit{Software:} \textit{Chempy} \citep{chempy}, \texttt{emcee} \citep{emcee}, \texttt{corner.py} \citep{corner}, \texttt{scikit-learn} \citep{scikit-learn}, \texttt{Keras} \citep{Keras}.

\bibliographystyle{apj}
\bibliography{exp_hyd}
\clearpage


\appendix
\begin{figure}[h]
\centering
\includegraphics[width=0.75\columnwidth]{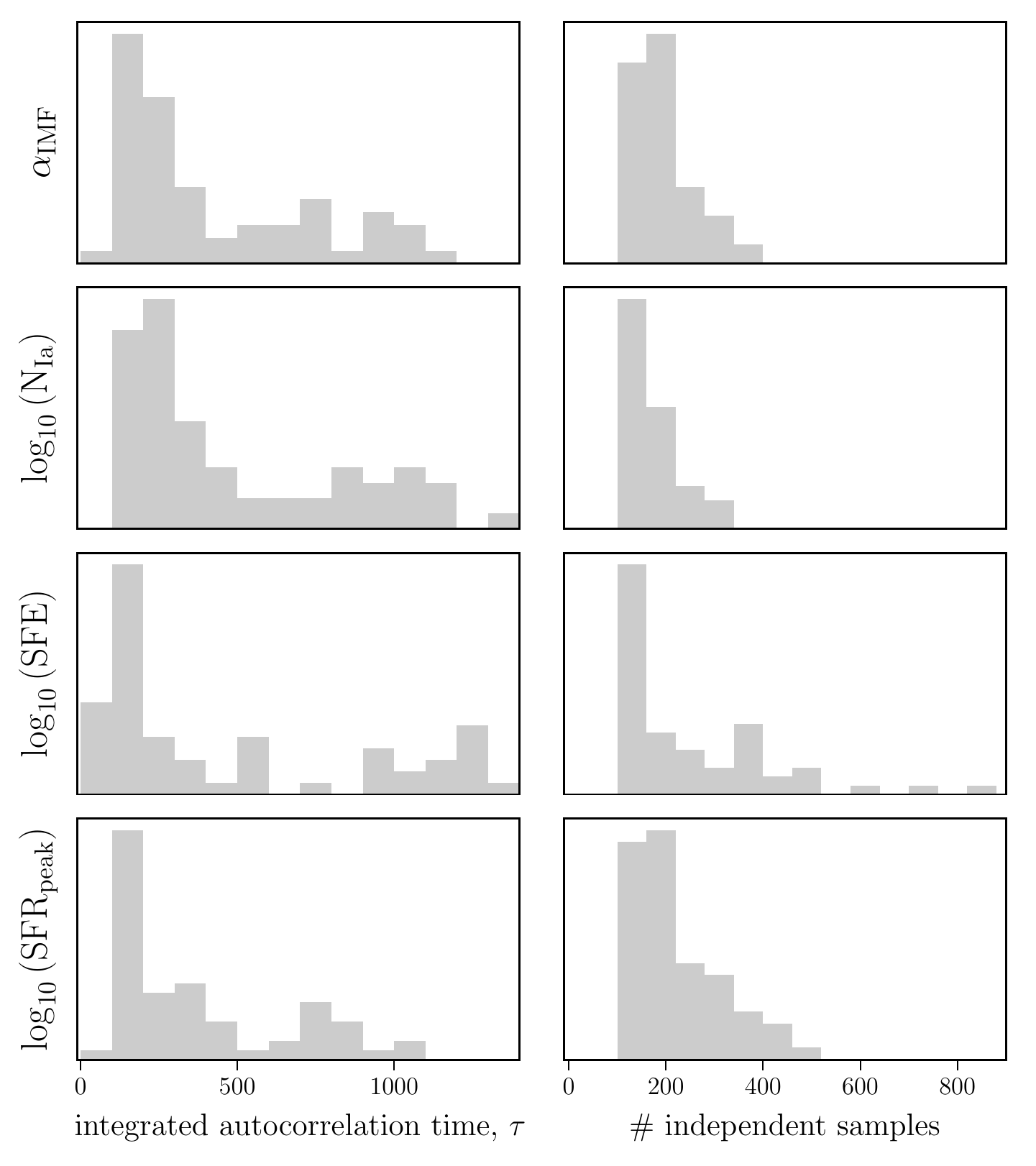}
\caption{Distribution of average integrated autocorrelation times (left), and the average number of estimated independent posterior samples (right) for each inferred \textit{Chempy} parameter.}
\label{fig:convergence}
\end{figure}

\begin{figure}[tp]
\centering
\includegraphics[width=0.8\columnwidth]{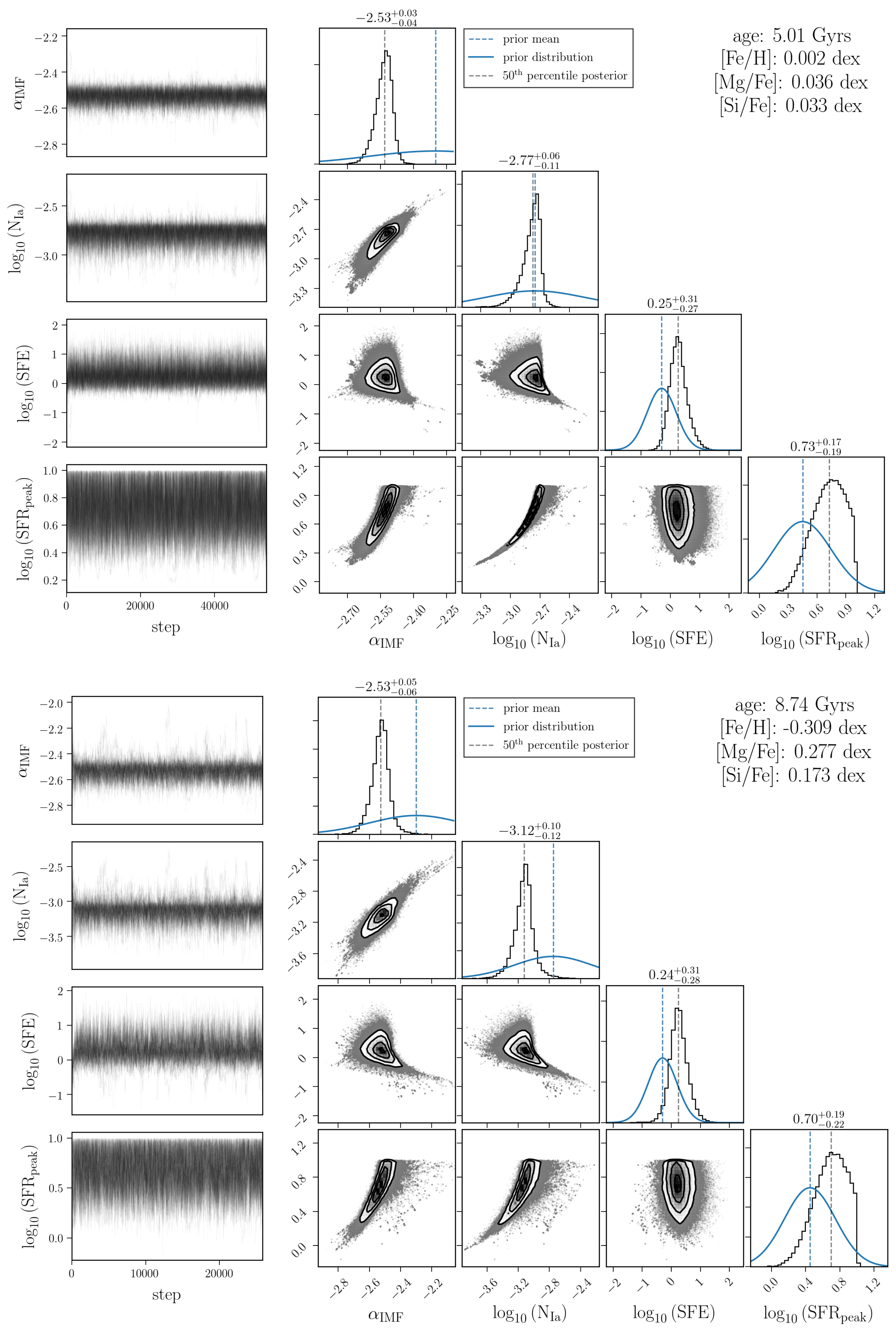}
\caption{Example trace plots (left) and joint posterior distributions (right) of the four inferred \textit{Chempy} parameters. The \textit{top panel} shows the posterior distributions for a solar-metallicity, $\sim$5 Gyr-aged star in the low-$\alpha$ sequence, and the \textit{bottom panel} shows the posterior distributions for a metal-poor, $\sim$9 Gyr-aged star in the high-$\alpha$ sequence. The blue curves indicates the prior for each parameter, while the histograms show the marginalized posterior distributions with the 50$^{\rm th}$ percentile values indicated by the dashed lines. The sharp edges in the SFR$_{\rm peak}$ posterior distributions are a result of the model being constrained to explore SFR peaks occurring before $\sim$12.6 Gyr.}
\label{fig:corner}
\end{figure}

\twocolumngrid

Figure \ref{fig:convergence} shows the distribution of the integrated autocorrelation times for each \textit{Chempy} sampled parameter. The average integrated autocorrelation time is computed across the chains for each star that we run. Also shown is the effective number of independent posterior samples, based on the estimated integrated autocorrelation times and chain lengths.

As an example, Figure \ref{fig:corner} shows the trace plots and joint posterior distributions for two stars that we fit with \textit{Chempy}: a solar-metallicity, $\sim$5 Gyr-aged star in the low-$\alpha$ sequence, and a metal-poor, $\sim$9 Gyr-aged star in the high-$\alpha$ sequence. As seen in the figure, the chains appear well-mixed and stationary. This, in addition to the integrated autocorrelation times, is suggestive of convergence. Examining the posteriors of the 5 Gyr-old star, we see that for the parameter describing the slope of the IMF, $\alpha_{\rm IMF}$, the posterior distribution peaks at lower values than the prior. This suggests that stellar populations with fewer high-mass to low-mass stars, compared to the typical Chabrier IMF, is necessary to produce the [Fe/H], [Mg/Fe], and [Si/Fe] abundances of this star. For the remaining parameters, the deviations between the posterior and prior distributions are less significant. The peak of the SN Ia parameter posterior is similar to the mean of the prior, but compared to the priors higher SFE and later SFR$_{\rm peak}$ values appear to be preferred. Considering the relationship between posterior distributions, the strongest correlations are between the IMF slope and number of SN Ia, the IMF slope and the SFR$_{\rm peak}$, and the number of SN Ia and the SFR$_{\rm peak}$. As described in \cite{chempy}, the correlation between the $\alpha_{\rm IMF}$ and N$_{\rm Ia}$ parameters is because an increased metal production from more CC-SN requires more Fe from SN Ia to generate the correct balance of $\alpha$-element and Fe enrichment. The SFR$_{\rm peak}$ is additionally correlated with $\alpha_{\rm IMF}$ and N$_{\rm Ia}$ parameters because a later SFR$_{\mathrm{peak}}$ increases the number of metal-rich stars, which eject more metals. 

The posteriors of the $\sim$9 Gyr-old metal-poor star exhibit behavior mostly similar to the posteriors of the $\sim$5 Gyr-old solar-metallicity star. The $\alpha_{\rm IMF}$ posterior for this star is nearly the same as that of the 5 Gyr-aged star, suggesting that the chemical abundance pattern of this star arose from a stellar population that also formed with fewer high-mass to low-mass stars, compared to the typical Chabrier IMF. The SFE and SFR$_{\rm peak}$ posteriors for this star are also similar to the 5 Gyr-aged star, with a higher efficiency and later SFR peak preferred compared to the priors. The main difference between the posteriors of the two stars is in the N$_{\rm Ia}$ parameter. For the metal-poor $\sim$9 Gyr-aged star, a case of fewer exploding SN Ia is preferred because this enables the production of the lower [Fe/H] and higher $\alpha$ abundances of this star. 

\label{lastpage}
\end{document}